\pdfoutput=1
\documentclass[%
 reprint,
 amsmath,amssymb,
 aps,
]{revtex4-1}

\usepackage{graphicx,txfonts}
\usepackage{chemarr}
\usepackage{times,longtable}
\usepackage{hyperref}
\usepackage{version}
\usepackage{color}
\usepackage{soul}

\usepackage{epsf,mathtools}
\usepackage[usenames,dvipsnames]{xcolor}

\graphicspath{{figures/},{figures/pdf/},{figures/eps/}}

\bibpunct{[}{]}{,}{n}{}{,}

\newcommand{\btil}{b^o}
\newcommand{\betil}{\beta_o}

\newcommand{\red}[1]{\textcolor{red}{ #1}}

\renewcommand{\hat}[1]{\widehat{#1}}

\newcommand{\ii}{\mathrm{i}}

\usepackage{chemarr}
\usepackage{epsf,mathtools}
\usepackage{graphicx}% Include figure files
\usepackage{dcolumn}% Align table columns on decimal point
\usepackage{bm}% bold math
\graphicspath{{figures/},{figures/pdf/},{figures/eps/}}

\begin{document}

%\preprint{APS/123-QED}

\title{RNA-based regulation: dynamics and response to perturbations of competing RNAs}% Force line breaks with \\

\author{Matteo Figliuzzi}
\affiliation{Dipartimento di Fisica, Sapienza Universit\`a di Roma, P.le A. Moro 2, 00185 Rome (Italy)}

\author{Andrea De Martino}
\thanks{Authors contributed equally}
\affiliation{Dipartimento di Fisica, Sapienza Universit\`a di Roma, P.le A. Moro 2, 00185 Rome (Italy)}
\affiliation{CNR-IPCF, Unit\`a di Roma, Rome (Italy)}
\affiliation{Center for Life Nano Science@Sapienza, Istituto Italiano di Tecnologia, Viale Regina Elena 291, 00161 Roma, Italy}

\author{Enzo Marinari}
\thanks{Authors contributed equally}
\affiliation{Dipartimento di Fisica, Sapienza Universit\`a di Roma, P.le A. Moro 2, 00185 Rome (Italy)}
\affiliation{Center for Life Nano Science@Sapienza, Istituto Italiano di Tecnologia, Viale Regina Elena 291, 00161 Roma, Italy}

%\date{~}% It is always \today, today,

\begin{abstract}
The observation that, through a titration mechanism, microRNAs (miRNAs) can act as mediators of effective interactions among their common targets (competing endogenous RNAs or ceRNAs) has brought forward the idea (`ceRNA hypothesis') that RNAs can regulate each other in extended `cross-talk' networks. Such an ability might play a major role in post-transcriptional regulation (PTR) in shaping a cell's protein repertoire. Recent work focusing on the emergent properties of the cross-talk networks has emphasized the high flexibility and selectivity that may be achieved at stationarity. On the other hand, dynamical aspects, possibly crucial on the relevant time scales, are far less clear. We have carried out a dynamical study of the ceRNA hypothesis on a model of PTR. Sensitivity analysis shows that ceRNA cross-talk is dynamically extended, i.e. it may take place on time scales shorter than those required to achieve stationairity even in cases where no cross-talk occurs in the steady state, and is possibly amplified. Besides, in case of large, transfection-like perturbations the system may develop strongly non-linear, threshold response. Finally, we show that the ceRNA effect provides a very efficient way for a cell to achieve fast positive shifts in the level of a ceRNA when necessary. These results indicate that competition for miRNAs may indeed provide an elementary mechanism to achieve system-level regulatory effects on the transcriptome over physiologically relevant time scales.
\end{abstract}

\pacs{Valid PACS appear here}% PACS, the Physics and Astronomy
\maketitle

\section{Introduction}

It is now well established that a large part of the eukaryotic transcriptome consists of non-coding RNAs, including numerous species (up to several hundreds in humans) of microRNAs (miRNAs) \citep{chekulaeva2009mechanisms}. miRNAs  play a central role in post-transcriptional regulation, as their protein-mediated binding to a messenger RNA (mRNA) results in either translational repression or mRNA degradation \citep{valencia2006control,bartel2004micrornas}. Their impact however might be much more far-reaching. On one hand, the involvement of miRNAs in peculiar motifs of the transcriptional regulatory network suggests that they could actively perform noise processing (most importantly, buffering) in gene expression \citep{tsang2007microrna,osella2011role}. On the other hand, by being able to target different mRNA species with different kinetics, they can in principle act as the mediators of an effective interaction between the mRNAs, such that a change in the transcription level of one mRNA can result in an alteration of the levels of another mRNA \citep{arvey2010target}. The so-called `ceRNA hypothesis' (whereby ceRNA stands for `competitive endogenous RNA') has attracted considerable attention lately \citep{salmena2011cerna}. According to it, in any given cell type, the protein repertoire is effectively influenced by the levels of the different miRNA species, in ways that depend (a) on the {\it a priori} possible couplings between miRNAs and mRNAs (the `miR program'), and (b) on the kinetics that governs the different interactions. In such a scenario, significant shifts in the protein composition of a cell can be obtained by altering the level of a small number of miRNA species. This mechanism is now believed to play an important role in many biological processes, from cell differentiation to cancer \citep{cesana2011long,karreth2011vivo,tay2011coding}. 

Placing the intuitive appeal of the ceRNA hypothesis on firm quantitative grounds is an important open challenge. In particular, one would like to understand which kinetic parameters control the emergence of the effective cross-talk between mRNAs and what type of effective interaction networks may result from such a simple titration mechanism. Recently, different theories have been proposed that attempt to answer these questions at steady state \citep{figliuzzi2013micrornas,ala2013integrated}. The key results of this kind of approaches lie, in our view, in the emergence of selectivity: at any given level of miRNAs, only a (potentially) small number of effective couplings between mRNAs can be active, and by changing the levels of miRNAs the structure of the network of effective couplings can be modified. This confers miRNA-mediated regulation remarkable flexibility and regulatory power. Still, questions about the validity of the steady-state assumption arise, as it is well known that, for instance, in processes like cell differentiation molecular levels are not stationary. It is therefore very important to understand (i) what are the typical timescales over which the steady-state scenario is established, (ii) whether steady-state like phenomenology may be observed during transients, and (iii) which kinetic parameters control timescales and responses away from the steady state.

In order to gain a quantitative understanding of these issues, we extend here the study of the model of post-transcriptional regulation introduced in \citep{figliuzzi2013micrornas} by characterizing the transient response of the system to perturbations, i.e. to changes in the RNA transcription rates. 

In the first part of this work, we focus on small perturbations. By analyzing (in Fourier space) the linearized dynamics of a system of $N$ ceRNAs jointly targeted by a single miRNA species, we recover the cross-talk scenario obtained in \citep{figliuzzi2013micrornas} for the steady-state, according to which effective interactions may occur only when miRNAs are only partially recycled upon complex degradation (or, in other words, when the rate of stoichiometric decay of the miRNA-RNA complex is non-zero). In addition, however, we show that a significant response may dynamically occur over finite time scales even when the rate of stoichiometric complex decay is zero and miRNAs are fully recycled (i.e. when the complex degradation channel is purely catalytic). This scenario is further studied in the important limiting cases in which complex dissociation is much faster, or much slower, than complex degradation, where the relevant timescales can be characterized in detail. 

The second part of this work focuses instead on large perturbations. By numerical analysis and analytical estimations we characterize the emergence of non-linear response. In specific, we uncover an ``extended'' type of cross-talk (not described by linear response theory) that is activated when perturbations overcome a given threshold. We'll argue that this regime may indeed be realized in experiments. Finally, relaxation times will be fully characterized.

In summary, we provide an overall dynamical characterization of the `ceRNA hypothesis' in the limit in which the dynamics can be described by mass-action kinetics and molecular noise can be neglected. The work is organized as follows: the case of small perturbations is dealt with in Section II, while Section III presents the analysis of the response to large perturbations and Section IV contains a discussion of results. Auxiliary results are detailed in the Supporting Text.

\section{Small perturbations}

\subsection{The model and its linearized dynamics}

Our starting point is the model defined in \cite{figliuzzi2013micrornas}. We consider a system with one miRNA species ($\mu$) and $N$ ceRNA species ($m_i$, $i=1,\ldots, N$) that can form $N$ species of complexes ($c_i$) with the miRNA, with the allowed processes
\begin{gather*}\label{processes}
\emptyset \xrightleftharpoons[d_i]{b_i} m_i ~~~~~~~~~~~~
\emptyset \xrightleftharpoons[\delta]{\beta} \mu ~~~~~~~~~~~~
\mu+m_i \xrightleftharpoons[k_i^-]{k_i^+} c_i\\
c_i \xrightharpoonup{\sigma_i} \emptyset ~~~~~~~~~~~~
c_i \xrightharpoonup{\kappa_i} \mu 
\end{gather*}
Arrow superscripts and subscripts denote the corresponding rates. Note that complex decay can occur both with (rate $\kappa_i$, catalytic channel) and without (rate $\sigma_i$, stoichiometric channel) miRNA recycling. The mass-action rate equations for the above system are given by
\begin{gather}
\frac{d}{dt} m_i =-d_i m_i + b_i-k_i^+ \mu m_i+k_i^- c_i\nonumber\\
\frac{d}{dt} \mu=-\delta \mu + \beta-\sum_i k_i^+ \mu m_i+\sum_i(k_i^-+\kappa_i) c_i \label{rateeqs}\\
\frac{d}{dt} c_i=-(\sigma_i+k_i^-+\kappa_i)c_i + k_i^+ \mu m_i \nonumber~~,
\end{gather}
where $m_i\equiv m_i(t)$ denotes the level of species $m_i$ (and similarly for $\mu$ and $c_i$). In the steady state, one can most notably characterize the `susceptibilities'
\begin{equation}\label{due}
\chi_{ij}^{ss}=\frac{\partial [m_i]}{\partial b_j}~~~~~(i\neq j)
\end{equation}
(where $[a]$ denotes the long-time limit of $a(t)$) as functions of $[\mu]$ and of the kinetic parameters. Generically, larger $\chi_{ij}^{ss}$s imply a larger effective cross-talk interaction between ceRNA $i$ and ceRNA $j$. Because the interaction is mediated by the miRNA, one may expect that much will depend on whether ceRNAs $i$ and $j$ are completely repressed (`bound' for short), completely  unrepressed (`free' for short), or partially repressed (`susceptible' for short) by the miRNA. In rough terms, in the `bound' case most ceRNAs are bound in complexes and hence are unavailable for translation, while in the `free' case the fraction of ceRNAs bound in complexes is small. In these regimes, the response of a ceRNA level to changes in the miRNA level is typically very small. On the other hand, in the `susceptible' regime ceRNA levels depend sensibly on the miRNA level. The emergent features of cross-talk at steady state are the following (see \cite{figliuzzi2013micrornas} for details):
\begin{enumerate}
\item[a.] selectivity: kinetic parameters can be tuned so as to couple only a subset of ceRNAs by strong cross-talk interactions; 
\item[b.] directionality: in presence of kinetic heterogeneities, $\chi_{ij}^{ss}\neq \chi_{ji}^{ss}$;
\item[c.] flexibility: the topology of the effective interaction pattern among ceRNAs depends on $[\mu]$;
\item[d.] relevance of stoichiometric processing: $\sigma_j=0$ implies $\chi_{ij}^{ss}=0$. However, the quantity
\begin{equation}\label{sss}
\widetilde{\chi}_{ij}^{ss}\equiv\lim_{\substack{\sigma_j \to\kappa_j\\ \kappa_j \to 0}}\frac{\sigma_j+\kappa_j}{\sigma_j}~\chi_{ij}^{ss}~~,%(\sigma_j)
\end{equation}
corresponding to the steady state susceptibility of a system without recycling for $\sigma_j=0$, may remain finite. 
\end{enumerate}
We shall now focus on the return to the steasy state following a small perturbation away from it. Let 
\begin{gather}
x_i(t)\equiv m_i(t)-[m_i] \nonumber\\
y(t)\equiv \mu(t)-[\mu]\\
z_i(t)\equiv c_i(t)-[c_i] ~~.\nonumber
\end{gather}
Upon linearizing (\ref{rateeqs}), the above variables are seen to obey the equations
\begin{gather}
\frac{d}{dt}x_i=-d_ix_i + \btil_i-k_i^+([ \mu]x_i+[ m_i]y)+k_i^-z_i\nonumber\\
\frac{d}{dt}y=- \delta y + \betil-\sum_i k_i^+([\mu]x_i+[ m_i]y)+\sum_i(k_i^-+\kappa_i)z_i \label{lineqs} \\
\frac{d}{dt}z_i=-(\sigma_i+k_i^-+\kappa_i)z_i + k_i^+([ \mu]x_i+[ m_i]y)\nonumber~~,
\end{gather}
where we have introduced (small) time-dependent variations of the transcription rates, i.e. $b_i~\to~b_i(t)=b_i+\btil_i(t)$ and  $\beta~\to~\beta(t)=\beta+\betil(t)$. In what follows we shall focus on the emergent behaviour of (\ref{lineqs}).

\subsection{Dynamical response}

Details of the analysis of (\ref{lineqs}) are reported in the Supporting Text.  It turns out that, in Fourier space (where $\hat{a}(\omega)$ denotes the Fourier transform of $a(t)$), response may be quantified through the dynamical susceptibility
\begin{equation}\label{susk1}
\hat{\chi_{ij}}(\omega)=\frac{\partial \hat{x_i}}{\partial \hat{\btil_j}}=
\begin{cases}
\Psi_{ij}(\omega)\chi_{ij}^{ss}%=\Psi_{ij}(\omega)\chi_{ij}^{SS}
& \text{if $\sigma_j\neq 0$}\\
\Phi_{ij}(\omega)\widetilde{\chi}_{ij}^{ss}%=\Phi_{ij}(\omega)\widetilde{\chi}_{ij}^{SS}
 &\text{if $\sigma_j= 0$}
\end{cases}
\end{equation}
where we have isolated the frequency-dependent part of the dynamical susceptibility in the functions $\Psi_{ij}$ and $\Phi_{ij}$. These functions can be factorized as the product of different filters
\begin{eqnarray}\label{filters}
\Psi_ {ij}(\omega)=S_j(\omega) D(\omega) J_i(\omega)J_j(\omega)\\
\Phi_{ij}(\omega)=C_j(\omega) D(\omega)J_i(\omega)J_j(\omega) \label{filt2}~~,
\end{eqnarray}
where 
\begin{gather}
J_i(\omega)=\frac{1+\rho_i}{1+\rho_i \frac{(1+\ii\omega\tau_{1,i})(1+\ii\omega\tau_{2,i})}{1+\ii\omega\tau_{3,i}}}\\
S_i(\omega)=\frac{1+\ii\omega\tau_{4,i}}{1+\ii\omega\tau_{3,i}}\\
C_i(\omega)=\frac{\ii\omega\tau_{5,i}}{1+\ii\omega\tau_{5,i}}\\
D(\omega)=\frac{1}{\tau_0\,\chi_{\mu\mu}^{ss}\,\Big[(\ii\omega \tau_0+1)+\sum_i \gamma_i J_i(\omega)\Big]}
\end{gather}
and we have introduced the time scales
\begin{gather}
\tau_0=\delta^{-1} \quad, \quad\tau_{1,i}=d_i^{-1}\quad ,\quad \tau_{2,i}=(\sigma_i+\kappa_i+k_i^-)^{-1} \nonumber\\
\tau_{3,i}=(\sigma_i+\kappa_i)^{-1} \quad ,\quad \tau_{4,i}=\sigma_i^{-1}~~, \quad \tau_{5,i}=\kappa_i^{-1}
\end{gather}
as well as the parameters
\begin{equation}
%F_i([\mu])=\frac{\mu_{0,i}}{[\mu]+\mu_{0,i}}~~~~~,~~~~~
\rho_i=\frac{\mu_{0,i}}{[\mu]} ~~~~~,~~~~~ \gamma_i=\frac{k_i^+ [m_i]}{\tau_{0}(1+\rho_i)}~~~~~,~~~~~
\mu_{0,i}=\frac{d_i}{k_i^+}(1+\phi_i)
\end{equation}
(note the key role that the latter parameters plays in \citep{figliuzzi2013micrornas}). The quantity $\chi_{\mu\mu}^{ss}$ represents instead the steady-state susceptibility of the miRNA level to (small) changes in its transcription rate.

Before discussing the behaviour of the filters (especially $J_i$ and $D$) and giving a physical interpretation, let us clarify the meaning of the different time scales. $\tau_0$ and $\tau_{1,i}$ represent, respectively, the (average) lifetime of the miRNA and of ceRNA $i$ in absence of interactions. $\tau_{2,i}$, $\tau_{3,i}$, and $\tau_{4,i}$ are instead related to the processing of complex $c_i$: $\tau_{2,i}$ is the average lifetime of complex $c_i$ before unbinding or being degraded;  $\tau_{3,i}$ represents the average time needed for complex $c_i$ to be degraded (in absence of unbinding); finally, $\tau_{4,i}$ and $\tau_{5,i}$ are, respectively, the average times required for complex $c_i$ to be degraded stoichiometrically and, respectively, catalytically (in absence of all other processes). Note that $\tau_{2,i}\leq\tau_{3,i}\leq\tau_{4,i}$, whereas $\tau_{3,i}=\tau_{5,i}$ if $\sigma_i=0$.

Concerning the filters, we begin by noting that (see Supporting Text) $J_i$ measures the inertia of ceRNA $i$ in responding to a change in the level the miRNA. In particular, it is related to the `gain'
\begin{equation}
g_{i\mu}(\omega)\equiv \frac{\partial \hat{x_i}}{\partial \hat{y}}
\end{equation}
by $J_i(\omega)=g_{i\mu}(\omega)/g_{i\mu}(0)$, so that when $J_i\simeq 1$ ceRNA $i$ is istantaneously at equilibrium with the miRNA. Furthermore, we can re-write it as
\begin{equation}
J_i(\omega)=\Bigg( \frac{1+\rho_i}{1+\rho_i B_i(\omega)}\Bigg)\frac{1}{1+\ii\omega\tau_{1,i}^{{\rm eff}}(\omega)}
\end{equation}
where
\begin{equation}
\tau_{1,i}^{{\rm eff}}(\omega)= \frac{B_i(\omega)\rho_i}{1+B_i(\omega)\rho_i}\,\tau_{1,i}
\end{equation}
with
\begin{equation}
B_i(\omega)=\frac{1+\ii\omega\tau_{2,i}}{1+\ii\omega\tau_{3,i}}~~~~~,~~~~~|B_i(\omega)|\leq 1~~. 
\end{equation}
%The presence of the frequency-dependent factors $B_i$ implies that deviations of $J_i$ from the behavior of a simple low pass filter may be transiently possible. 
As shown in \citep{figliuzzi2013micrornas}, ceRNA $i$ is in the `bound' regime at steady state when $\rho_i\ll 1$ (i.e. when the miRNA level is much larger than a ceRNA-dependent threshold), while it is `free' and hence fully available for translation when $\rho_i\gg 1$. In these limits, the above expression for $J_i$ simplifies as
\begin{equation}\label{rhobound}
J_i(\omega)\simeq 
\begin{cases}
[1+\ii \omega\rho_i B_i(\omega)\tau_{1,i}]^{-1} &\text{for $\rho_i\ll1$}\\
B_i(\omega)^{-1}(1+\ii\omega\tau_{1,i})^{-1} &\text{for $\rho_i\gg 1$}
\end{cases}~~.
\end{equation}
From $|B_i(\omega)|\leq 1$, it follows that $J_i$ can be larger than one, implying the possibility that the ceRNA response to a variation of the miRNA level is transiently amplified with respect to corresponding the steady-state response.

Filter $D$ is common to all pairs of ceRNAs, and is strongly dependent on the miRNA decay timescale $\tau_0$. Note that $D$ and $J_i$'s are low pass filters, $C_i$'s are high-pass filters, while $S_i$'s allow for transmissions at any frequency (but preferentially transmit high frequencies), since
\begin{gather}
\lim_{\omega \to 0} J_i(\omega)=1 \quad,\quad \lim_{\omega \to \infty} J_i(\omega)=0\\
 \lim_{\omega \to 0} C_i(\omega)=0 \quad,\quad \lim_{\omega \to \infty} C_i(\omega)=1 \\
\lim_{\omega \to 0} S_i(\omega)=1 \quad,\quad
\lim_{\omega \to \infty} S_i(\omega)=\frac{\sigma_i+\kappa_i}{\sigma_i}\geq 1\\
\lim_{\omega \to 0} D(\omega)=1 \quad , \quad \lim_{\omega \to \infty} D(\omega)=0 ~~.
\end{gather}
The complete absence of stoichiometric processing (and hence full miRNA recycling) strongly affects the dynamical behaviour of the system: indeed different filters ($C_i$ for completly catalytic, $S_i$ for at least partially stoichiometric processing) describe the two situations.

One can now see that the steady-state crosstalk scenario is recovered the limit $\omega\to 0$. Indeed, because $C_j(0)=0$, in absence of stoichiometric decay ($\sigma_j=0$) one has $\hat{\chi_{ij}}(0)=0$: at steady state, cross-talk is possible only when $\sigma_j\neq 0$, in agreement with \citep{figliuzzi2013micrornas}. Away from the steady state, however, the situation changes. In particular, the dynamical susceptibility (\ref{susk1}) in case of completely catalytic degradation ($\sigma_j=0$) contains both low- and high-pass filters. As a consequence, we expect that in this case it will be possible for the system to transmit a signal at intermediate frequencies, i.e. to observe a response on intermediate time scales.

\subsection{Timescale separation: limiting cases}

Unfortunately, little is known about the kinetics of RNA interference. Studies on association kinetics between RNAs in prokaryotic systems indicate that complexes (formed, in that case, by mRNAs and small RNAs) might dissociate into their original components with rates $k_i^-\simeq 10^{-2} \div 10^{-1}$ s$^{-1}$ higher than the processing rates $\sigma_i$ and $\kappa_i$ of the complex \citep{argaman2000fhla,wagner200212}. On the other hand, analysis of RNA interference in eukaryotes suggest that the kinetics can vary substantially across different targets and that it is strongly affected by the degree of complementarity \citep{haley2004kinetic}. In absence of more precise information, we shall focus on the limiting behaviour in two cases, namely those of ``slow'' ($k_i^-\ll \kappa_i+\sigma_i$) and ``fast'' ($k_i^-\gg \kappa_i+\sigma_i$) complex dissociation. The remaining parameters used in the following numerical studies are set are to biologically reasonable values falling inside the ranges considered in \citep{haley2004kinetic,alon2007introduction,wang2010toward} and are measured in units of typical RNA half life ($\tau\simeq 10^4$ s) and typical RNA concentration ($\gamma\simeq 1$ nM). In these units, $d_i\simeq\delta\simeq\tau^{-1}$, $b_i\simeq \gamma \tau^{-1}$, $k_i^+\simeq 1 \div 100\, \gamma^{-1} \tau^{-1}$, and $\kappa_i \simeq 1 \div 10 \,\tau^{-1}$.% and, as stated before, we will focus on the catalytic case: $\sigma_i=0$.

\subsubsection{Slow complex dissociation}

In this limit, complex formation is far from equilibrium at steady state (in favor of association). Moreover, $\tau_{2,i}\simeq \tau_{3,i}$, so that $B_i(\omega)\simeq 1$ at any frequency. %\footnote{first order correction: $B_i(\omega)\approx 1-\frac{k_i^-}{\sigma_i+\kappa_i}\Big(\frac{i\omega \tau_3}{1+i\omega\tau_3}\Big)$}. 
Therefore, $J_i(\omega)$ behaves as a simple low pass filter, i.e.
\begin{equation}
J_i(\omega)\simeq\frac{1}{1+\ii\omega \tau_{1,i}^{{\rm eff}}}~~,
\end{equation}
where
\begin{equation}
\tau_{1,i}^{{\rm eff}}\simeq\frac{\rho_i}{1+\rho_i}\tau_{1,i}~~.
\end{equation}
Note that 
\begin{equation}
\tau_{1,i}^{{\rm eff}}\,\,
\begin{cases}
\simeq \tau_{1,i} & \text{if $\rho_i\gg 1$}\\ 
\simeq \tau_{1,i}/2 & \text{if $\rho_i\simeq 1$}\\
\ll \tau_{1,i} & \text{if $\rho_i\ll 1$} 
\end{cases} ~~.
\end{equation}
Hence the cutoff frequency depends on $\rho_i$ and it is shifted towards high frequencies when ceRNAs are `bound' ($\rho_i\ll 1$). Following \citep{figliuzzi2013micrornas}, we will call `susceptible' the ceRNAs such that $\rho_i\simeq 1$.

\begin{figure}
\centering
{\includegraphics[width=0.48\textwidth]{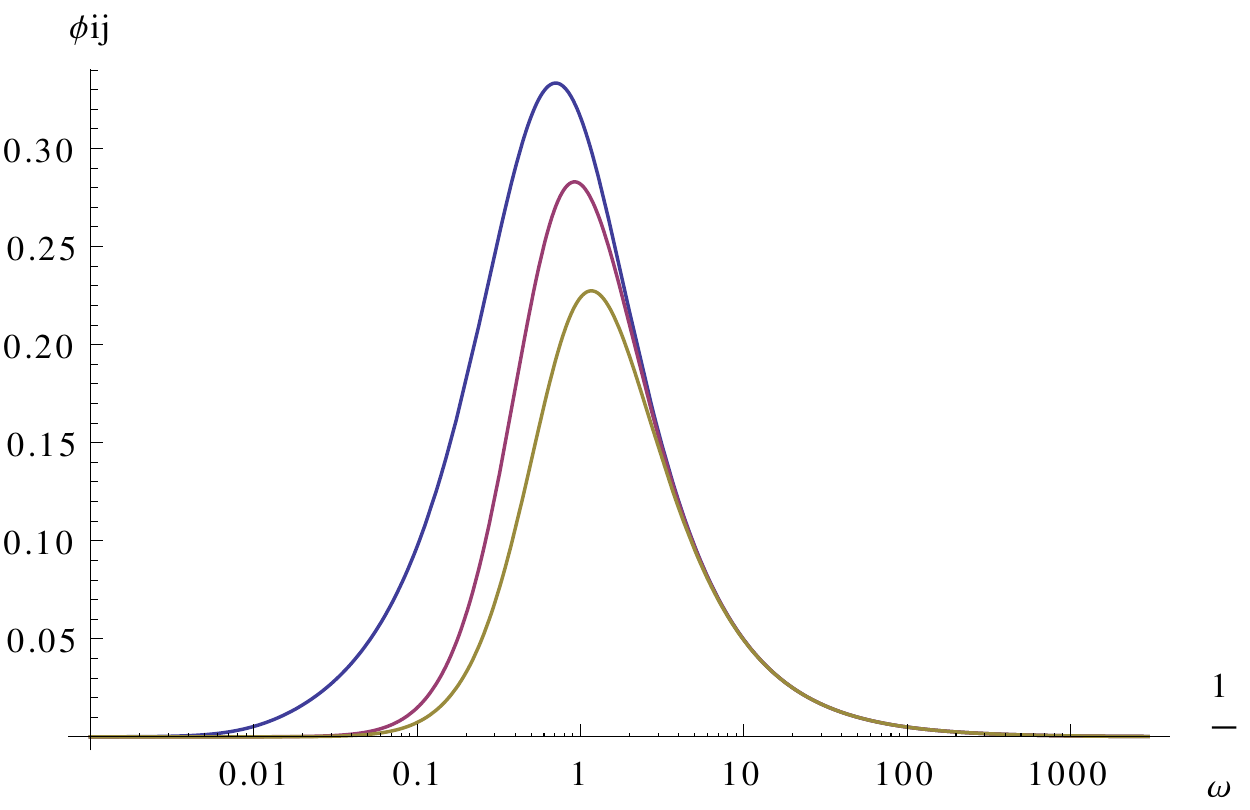}}\\
%{\includegraphics[width=0.45\textwidth]{slow_chi.pdf}}
\caption{ {\bf Slow dissociation, fast processing} Dynamical response for slow complex dissociation in a fully catalitic system ($\sigma_i=0,\kappa_i=10$) for pairs of `free' ($\rho_i=100$, in yellow), `susceptible' ($\rho_i=1$, in red) and `bound' ($\rho_i=0.01$, in blue) ceRNAs. Remaining parameters are set as follows: $d_i=1, k_i^-=0, \delta=1, Z_i\equiv(1+\ii\omega\tau_{1,i})(1+\ii\omega\tau_{2,i})/(1+\ii\omega\tau_{3,i})=10$ for each $i$. \label{fig1}}
\end{figure}

Figure \ref{fig1} shows the dynamical response for pairs of `free', `susceptible' and `bound' ceRNAs in the slow dissociation limit (we considered irreversible binding, i.e. $k^-_i=0 \quad \forall i$). The highest values of the global filter $\Phi_{ij}(\omega)=D(\omega)C_j(\omega)J_i(\omega)J_j(\omega)$ (see (\ref{filt2})) are achieved between pairs of `bound' ceRNAs, when $\omega\simeq 1$. 
%The low-pass filters $J_i(\omega)$ for Bound ceRNAs has a much lower cutoff ($\omega_{th} \sim \frac{1}{10}\tau^{(1)}$) respect to the case of Free or Susceptible ceRNAs ($\omega_{th} \sim \tau^{(1)}$): 
In Figure \ref{fig1s} of the Supporting Text the Susceptibility $\chi_{ij}(\omega)$ is shown: notice that at high frequencies cross-talk between `bound' ceRNAs is stronger than that between `free' ceRNAs.

\begin{figure}
\centering
{\includegraphics[width=0.48\textwidth]{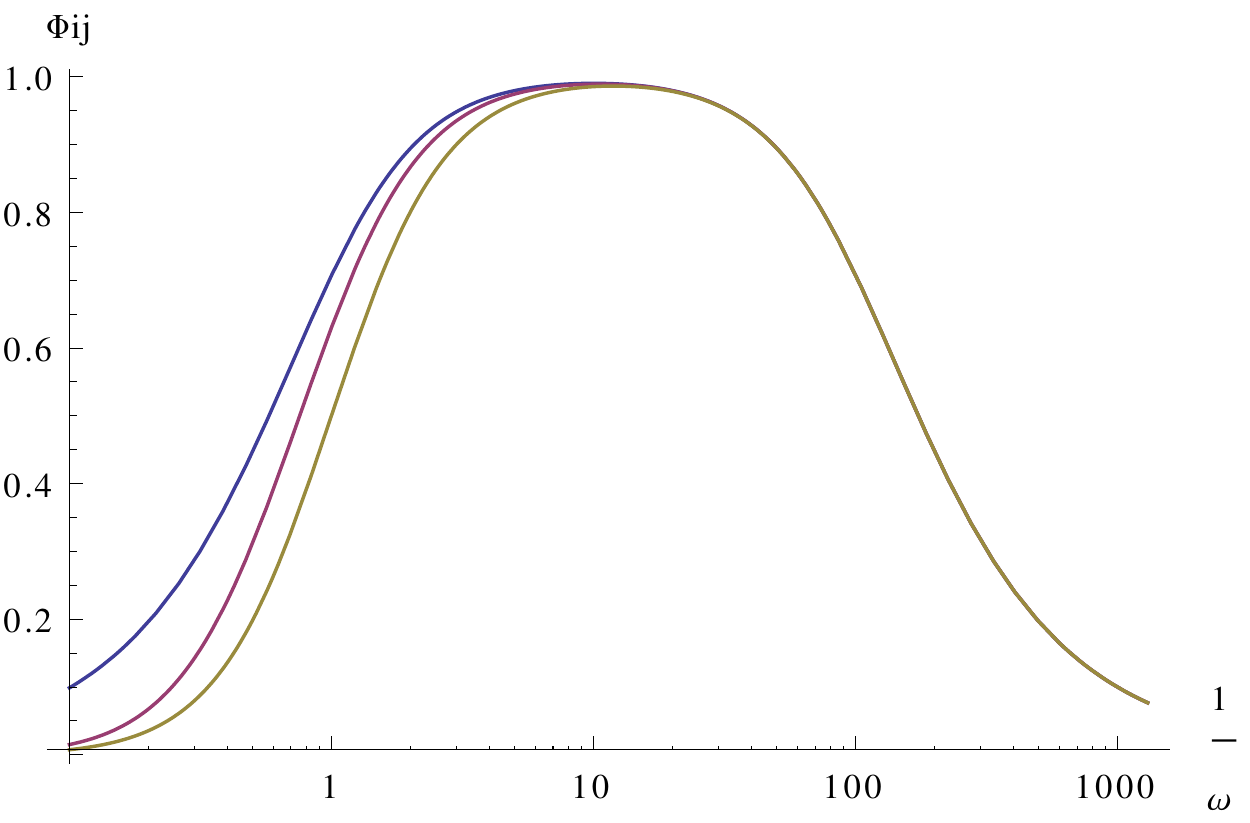}}\\
%{\includegraphics[width=0.45\textwidth]{recover_chi.pdf}}
\caption{ {\bf Slow dissociation, slow processing} Dynamical response in a fully catalitic system ($\sigma_i=0,\kappa_i=0.01$) for a couple of free ceRNA ($\rho_i=100$, in yellow), for a couple of susc ceRNA ($\rho_i=1$, in red), for a couple of bound ceRNA ($\rho_i=0.01$, in blue). Other parameters are set as follows: $d_i=1, k_i^-=0, \delta=1, Z_i=10$ for each $i$. \label{fig2}}
\end{figure}

If $\kappa_i<d_i$  we expect to recover a cross-talk scenario even for a fully catalytic system with $\sigma_i=0$ (for which no cross-talk may occur at steady state). Indeed
\begin{equation}
\hat{\chi_{ij}}(\omega)\simeq 
\begin{cases}
\widetilde{\chi}_{ij}^{ss}/(\omega^2\tau_{1,i}\tau_{2,i}) & \text{if $\omega\gg d_j$}\\
\widetilde{\chi}_{ij}^{ss} & \text{if $\kappa_j \ll \omega\ll d_j$}\\
-\ii\omega\widetilde{\chi}_{ij}^{ss}/\kappa_j & \text{if $\omega\ll \kappa_j$}
\end{cases}~~.
\end{equation}
Figure \ref{fig2} shows that in case of slow catalytic processing we recover the stoichiometric steady-state scenario for intermediate frequencies: in the frequency window between $\omega\simeq d_i=1$ and  $\omega\simeq \kappa_i$ the global filter $\Phi_{ij}$ is close to $1$. Correspondingly, $\chi_{ij}(\omega)\approx \tilde{\chi}_{ij}$, as shown in Figure \ref{fig2s} of the Supporting Text.

\subsubsection{Fast complex dissociation}

\begin{figure}
\centering
{\includegraphics[width=0.48\textwidth]{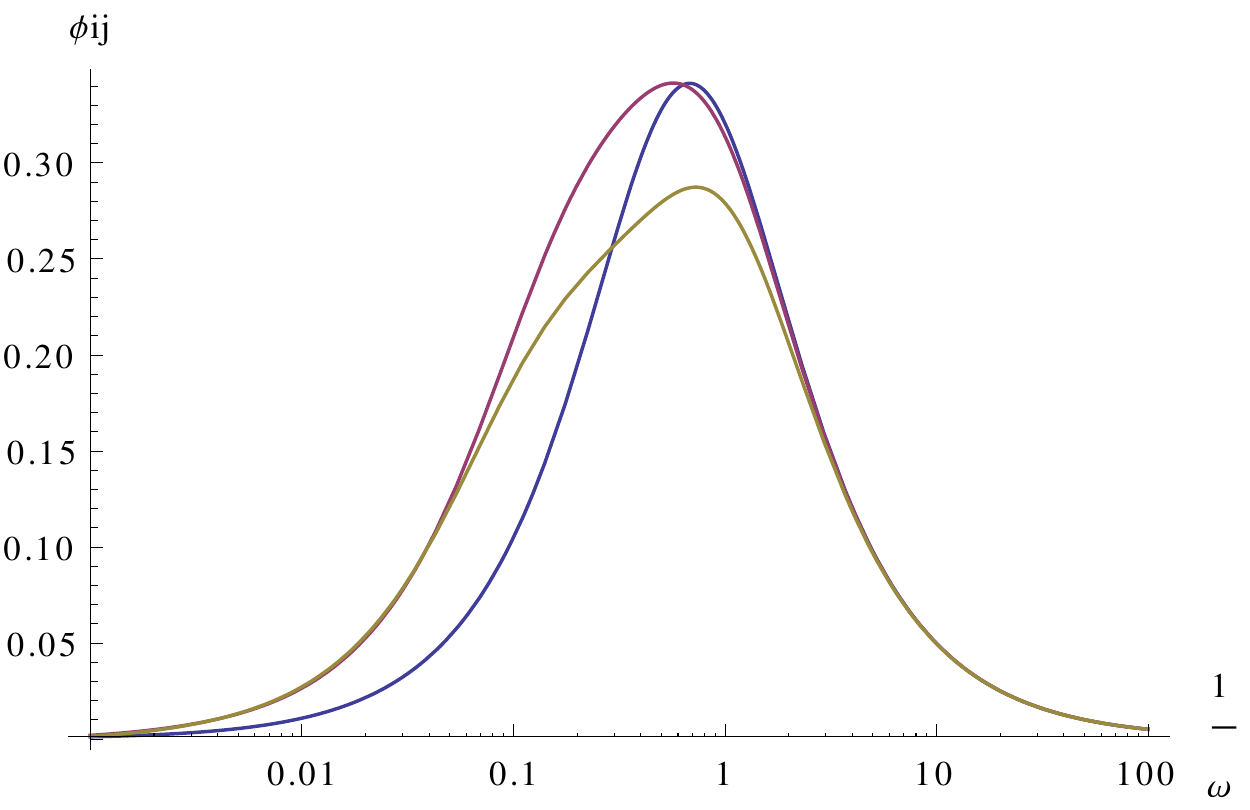}}\\
\caption{ {\bf Fast dissociation, slow processing} Dynamical response for fast complex dissociation in a fully catalitic system ($\sigma_i=0,\kappa_i=10$) for pairs of `free' ($\rho_i=100$, in yellow), `susceptible' ($\rho_i=1$, in red) and `bound' ceRNAs ($\rho_i=0.01$, in blue).  Other parameters are set as follows: $d_i=1, k_i^-=1000, \delta=1, Z_i=10$ for each $i$. \label{fig3}}
\end{figure}

In this case, the levels of complexes are close to equilibrium at steady state, while $\tau_{3,i}\gg\tau_{2,i}\simeq 1/k_i^-$, so that
\begin{equation}
   B_i(\omega)\simeq \frac{1+\ii\omega/k_i^-}{1+\ii\omega\tau_{3,i}}~~.   
\end{equation}
At low enough frequencies ($\omega\ll k_i^-$), $B_i(\omega)\simeq 1$ and we recover the `slow complex dissociation' scenario, while for high enough frequencies ($ \omega \gg \kappa_i+\sigma_i$), $B_i(\omega)%\simeq \frac{\kappa_i+\sigma_i}{k_i^-}
\ll 1$ and, as before, $J_i$ can be expressed as a simple low pass filter, viz.
\begin{equation}
J_i(\omega)\simeq\frac{1+\rho_i}{1+\alpha_i\rho_i}\frac{1}{1+\ii\omega \tau_{1,i}^{{\rm eff}}}
\qquad,\qquad \alpha_i=\frac{\kappa_i+\sigma_i}{k_i^-}~~,
\end{equation}
with 
\begin{equation}
\tau_{1,i}^{{\rm eff}}=\frac{\alpha_i\rho_i}{1+\alpha_i\rho_i}\tau_{1,i}~~.
\end{equation}
So we see that in this case we have both a regime-dependent cut-off frequency, as in the case of slow unbinding, and a regime-dependent modulation that amplifies ceRNAs cross-talk. Figure \ref{fig3} and Figure \ref{fig3s} of the Supporting Text show that indeed the dynamical response scenario for fast dissociation is similar to the one obtained for slow dissociation, except at high frequencies, where the cross-talk between `free' ceRNAs is stronger than the cross-talk between `bound' species.

\begin{figure}
\centering
{\includegraphics[width=0.48\textwidth]{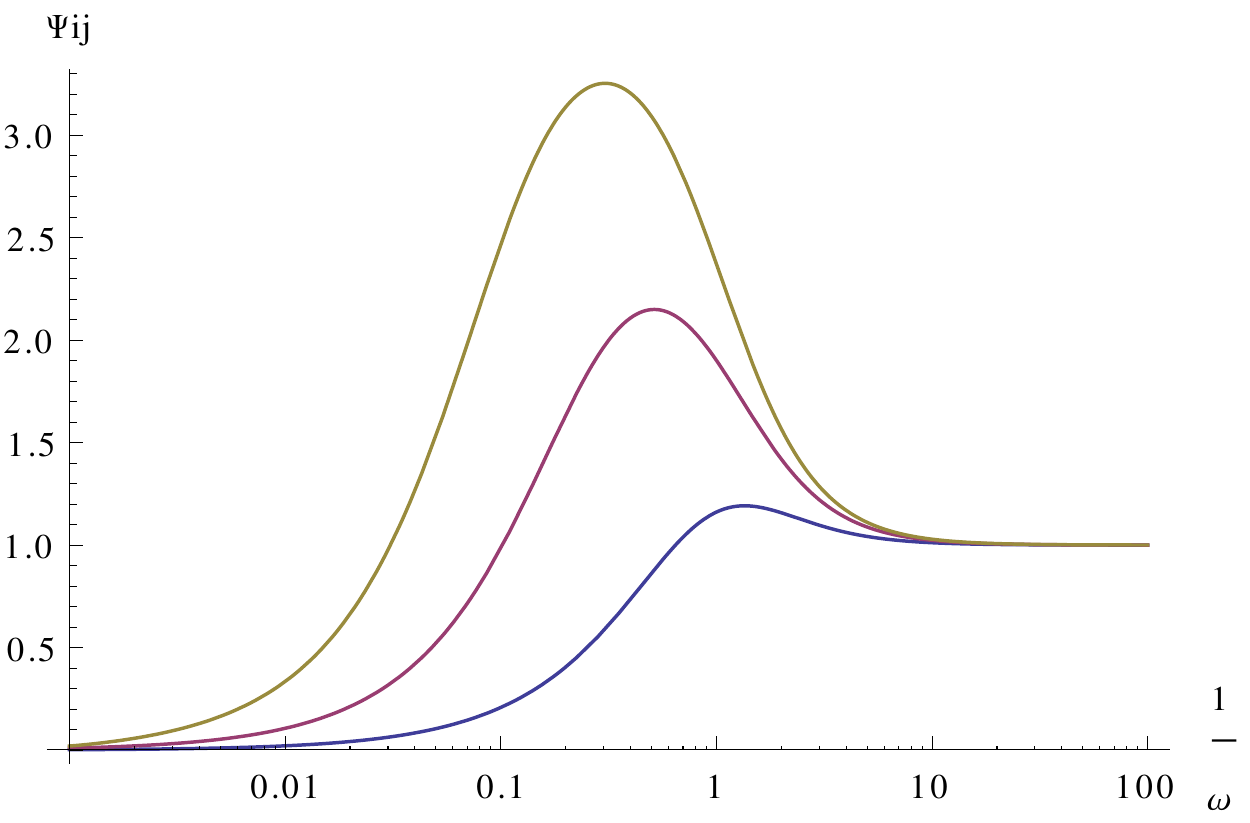}}\\
\caption{ {\bf Fast dissociation, fast processing} Dynamical response for fast complex dissociation in a fully stoichiometric system ($\sigma_i=0.5,\kappa_i=0$) for pairs of `free' ($\rho_i=100$, in yellow), `susceptible' ($\rho_i=1$, in red) and `bound' ($\rho_i=0.01$, in blue) ceRNAs. Other parameters are set as follows: $d_i=1, k_i^-=1000, \delta=1, Z_i=10$ for each $i$. \label{fig4}}
\end{figure}

Note that if the processing of complexes is slower than spontaneous degradation, i.e. if $\tau_{5,i}>\tau_{1,i}^{{\rm eff}}$, then cross-talk can be dynamically amplified. Indeed taking the expression (\ref{rhobound}) for `bound' ceRNAs ($\rho_i\ll1$) we observe that
\begin{equation}
\hat{\chi_{ij}}(\omega) \simeq \frac{1+\rho_i}{1+\alpha_i\rho_i}\frac{1+\rho_j}{1+\alpha_j\rho_j}\widetilde{\chi}_{ij}^{ss} 
\end{equation} 
which exceeds $\widetilde{\chi}_{ij}^{ss}$ for $\kappa_i<\omega<d_i/(\rho_i\alpha_i)$. 
 In order to have a direct comparison with the steady state scenario, we have considered the case of a fully stoichiometric system:
Figure \ref{fig4} shows  that, in case of slow processing (slower than spontaneous decay), dynamical cross-talk (quantified in this case by the filter $\Psi_{ij}(\omega)$) is stronger than the stoichiometric steady-state counterpart (i.e. $\Psi_{ij}(\omega)>1$), in particular for `susceptible' and `free' species.
Dynamical susceptibilities $\chi_{ij}(\omega)$ for the same choice of kinetic parameters are again shown in the Supporting Text (Figure \ref{fig4s}).

\section{Large perturbations}

When perturbations that bring the system away from the steady state are sufficiently large, deviations from the linear response scenario occur. The characterization of these phenomena are especially important to understand experiments, since transcriptional perturbations are normally carried out by transfections that increase levels by several folds, and it is on them that we shall focus in this section. In particular, we will show (numerically) that under large perturbations (a) the response can be highly non linear, (b) cross-talk is extended, in that it may take place between pairs of ceRNAs that would not interact otherwise, and (c) there exists a threshold perturbation for activating such an extended crosstalk. In addition, the response to large perturbations appears to be characterized by saturation effects as wells as by characteristic times increasing linearly with the perturbation.

We shall consider a particular kind of perturbation, namely a step-like transcriptional input that modifies the transcription rate of a given ceRNA at time $t=t_0$, defined by
\begin{equation}
b_i(t)= b_i\Big[1+\Delta_i\theta(t-t_0)\Big]
\end{equation}
where $\theta(x)$ is the Heavyside step function and $\Delta_i$ measures the fold change in the transcription rate of ceRNA $i$ after time $t_0$. We will focus the analysis on completely catalytic systems ($\sigma_i=0$ for all $i$), for which we have seen in the previous section that ceRNA crosstalk is activated for a finite time interval. In order to quantify the response of the system, we shall resort to an Integrated Response (IR) defined as
\begin{equation}\label{irr}
{\rm IR}_{ij}(\Delta_j)=\int_{t_0}^{\infty} \left[m_i(t+t_0)-m_i(t_0)\right] dt
\end{equation}
which depends both on the size of the response and on its duration. If only free mRNA molecules are translated into protein (at constant rate), (\ref{irr}) is strictly related to the total amount of protein produced in response to the perturbation, i.e. to the ultimate output of the input transcriptional signal. When perturbations are large, the time needed to relax back to the steady state after a transcriptional perturbation may be long compared to cellular processes and can vary according to the specific conditions \cite{morris2004small}. We will attempt, in case of large perturbation, to characterize such a relaxation time.

\subsection{Extended cross-talk}%: $\mathcal{B}-\mathcal{B}$ and $\mathcal{F}-\mathcal{F}$ interactions}

According to the linear response theory developed in \cite{figliuzzi2013micrornas}, cross-talk may take place only between `susceptible' ceRNAs (symmetrically, i.e. perturbing one species causes a response in the other and vice-versa) and from `bound' to `susceptible' ceRNAs (asymmetrically, i.e. perturbing a `bound' ceRNA a `susceptible' one will respond, but not vice-versa). We will see here that, when the perturbation overcomes a certain threshold, cross-talk is no longer limited to the above cases. 

\begin{figure}
\centering
{\includegraphics[width=0.48\textwidth]{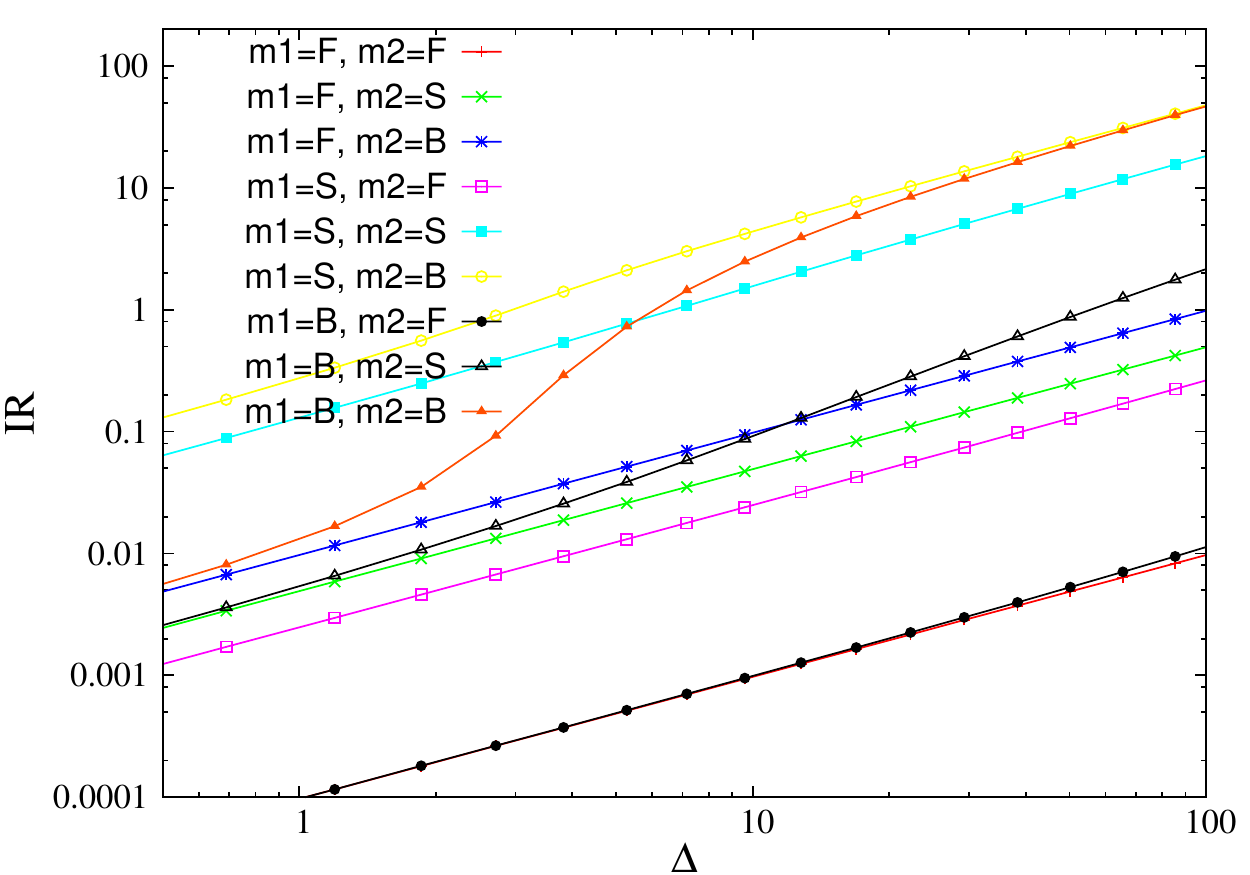}}
\caption{Integrated response as a function of the perturbation size.  ceRNAs have different binding rates so that they can be in the free (F), susceptible (S) or bound (B) regime, according to the legend. Binding rates are: $k^+=10^{-2}$ for free ceRNAs, $1$ for susceptible ceRNAs, and $10^2$ for bound ceRNAs. Remaining kinetic parameters are as follows: $b_1=b_2=\beta=1$, $d_1=d_2=\delta=1$, $k_1^-=k_2^-=0$, $\kappa_1=\kappa_2=1$.}
\label{IRup}
\end{figure}

\begin{figure}
\centering
\includegraphics[width=0.23\textwidth]{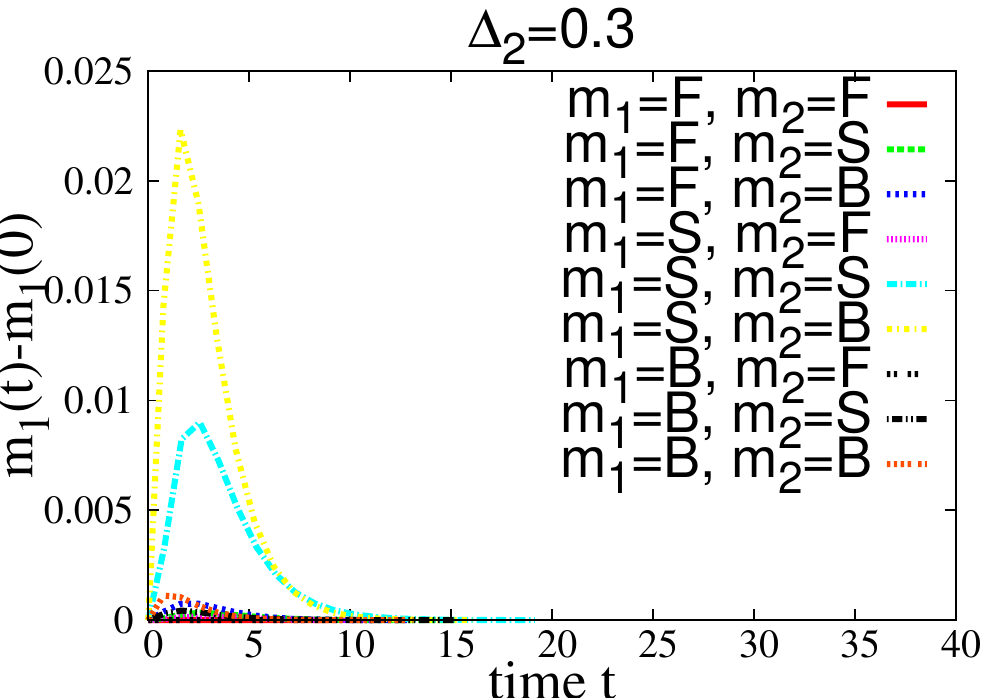}
%\caption{{\bf transcription perturbation:} time dependent response of cerna 1, with after the perturbation $\delta b_2=0.8$ at $t=0$($b_\mu=1=\mu^{SS}$) Cernas have $k^+=100,1,0.01$: they are in the bound, susc and free regime respectively ($\epsilon=0.01$)}
\includegraphics[width=0.23\textwidth]{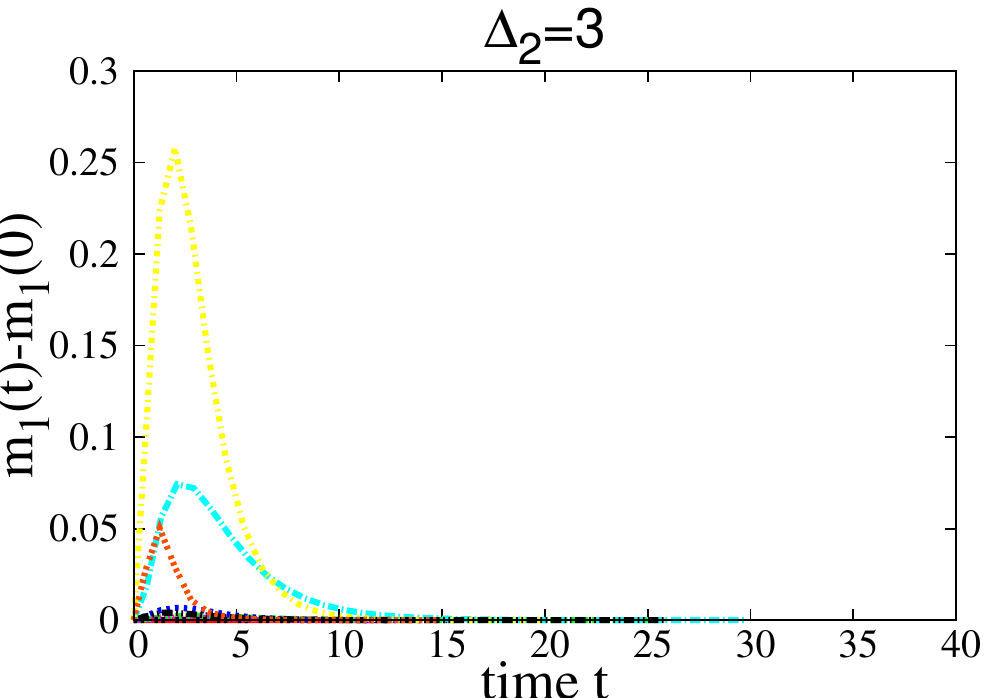}\\
%\caption{{\bf transcription perturbation:}time dependent response of cerna 1, with after the perturbation $\delta b_2=8.4$ at $t=0$($b_\mu=1=\mu^{SS}$)  }
\includegraphics[width=0.23\textwidth]{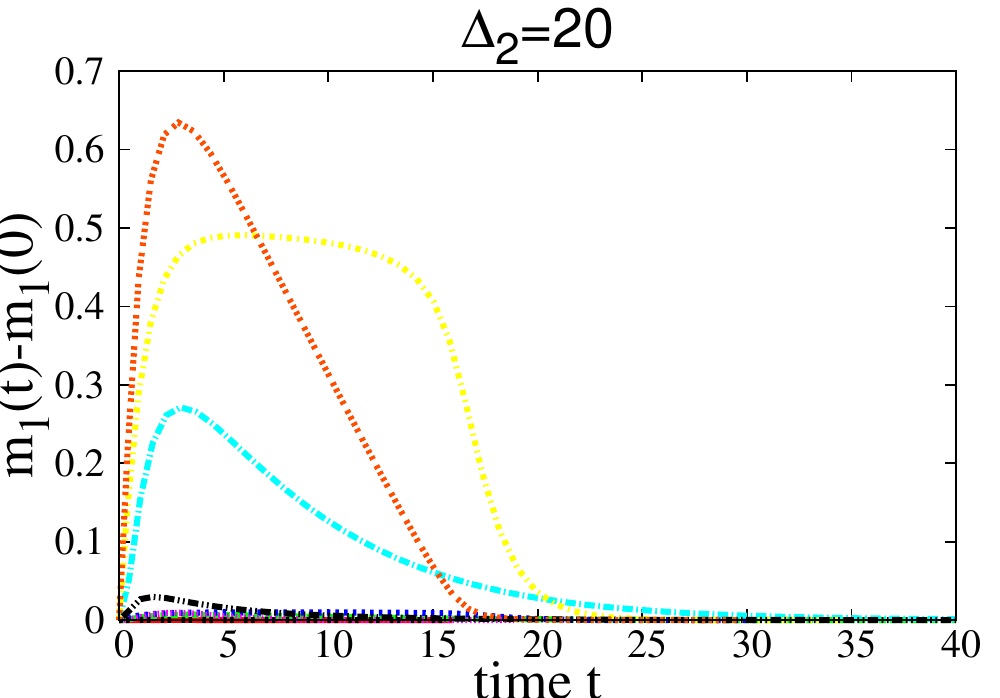}
%\caption{{\bf transcription perturbation:}time dependent response of cerna 1, with after the perturbation $\delta b_2=50$ at $t=0$($b_\mu=1=\mu^{SS}$)  }
\includegraphics[width=0.23\textwidth]{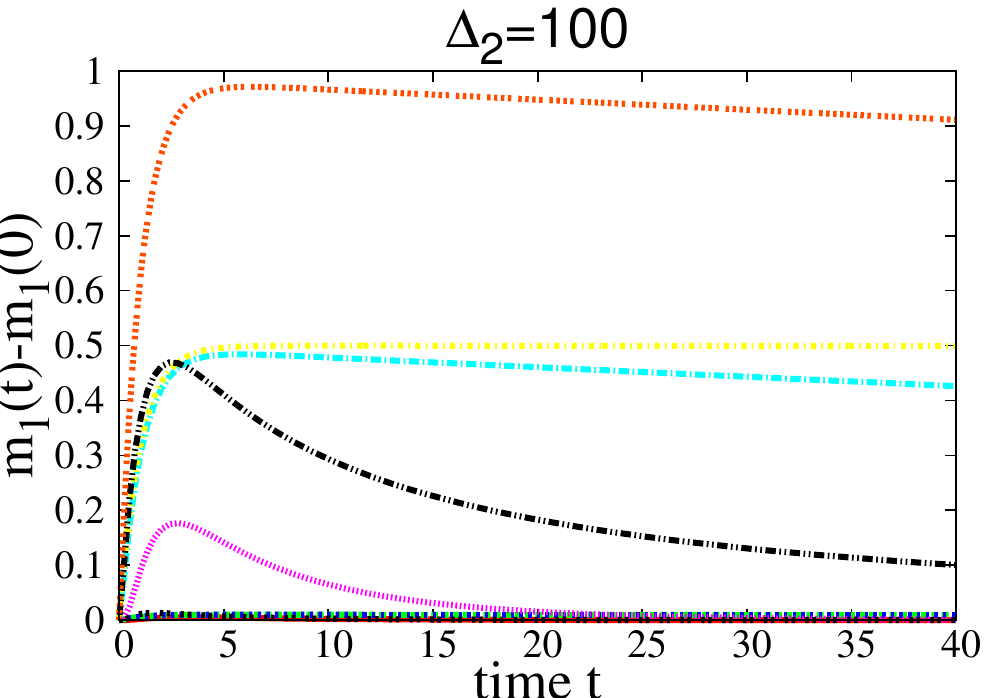}
%\caption{{\bf transcription perturbation:}time dependent response of cerna 1, with after the perturbation $\delta b_2=800$ at $t=0$($b_\mu=1=\mu^{SS}$)  }
\caption{Dynamical evolution of ceRNA 1 ($\Delta m_i(t)=m_1(t)-m_1(0)$) after a positive transcriptional perturbation $\Delta_2$ of ceRNA 2 at $t=0$. ceRNAs have different binding rates so that they can be in the free (F), susceptible (S) or bound (B) regime. Note that in the top plots only the B-S and S-S crosstalk is activated, while in the bottom ones (where the perturbation is above the threshold) the B-B coupling has switched on. 
}
\label{IRup2}
\end{figure}

Figure  \ref{IRup} shows the Integrated response of ceRNA1, after a positive perturbation $\Delta_2$ on the transcription rate of ceRNA2: strong deviations from linearity appear in the IR between bound ceRNAs, and if the perturbation is large enough the cross-talk between `bound' species can overcome that between `susceptible' ones. Indeed while integrated response is almost linear in the perturbation size for ceRNAs in the free and in the susceptible regimes, it is strongly non linear in the case of bound ceRNAs, specifically when the perturbation overcomes a given threshold (in this case $\Delta_2 \simeq 4$). The same effect is also evident in Figure \ref{IRup2}.

\begin{figure}
\centering
{\includegraphics[width=0.48\textwidth]{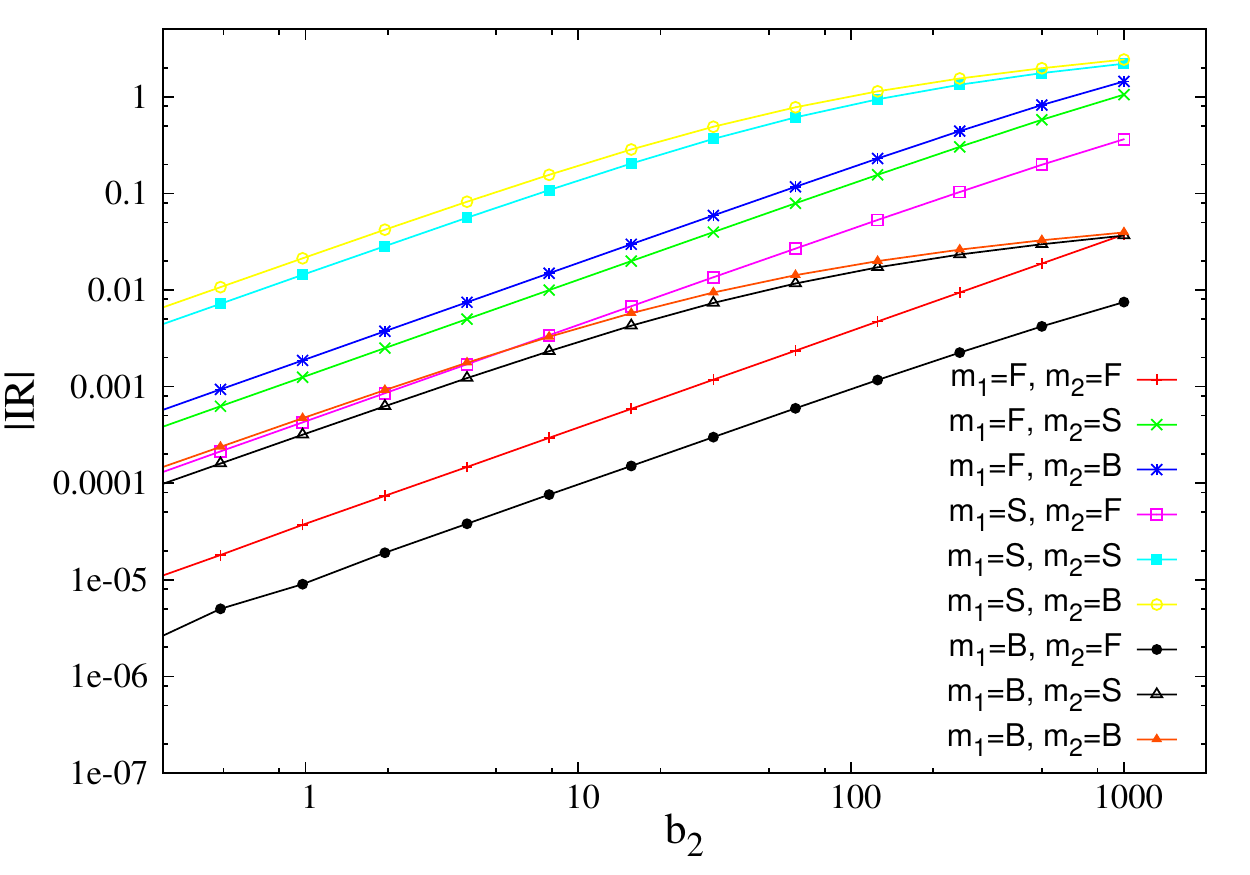}}
\caption{Absolute values of integrated response as a function of the perturbation size. ceRNAs have different binding rates so that they can be in the free (F), susceptible (S) or bound (B) regime. Binding rates are: $k^+=10^{-2}$ for free ceRNAs, $1$ susceptible ceRNAs, $10^2$ for bound ceRNAs. Remaining kinetic parameters are as follows: $b_1=\beta=1$, $d_1=d_2=1$, $\delta=0.5$, $k_1^-=k_2^-=0$, $\kappa_1=\kappa_2=1$.}
\label{IRdown}
\end{figure}

\begin{figure}
\centering
\includegraphics[width=0.23\textwidth]{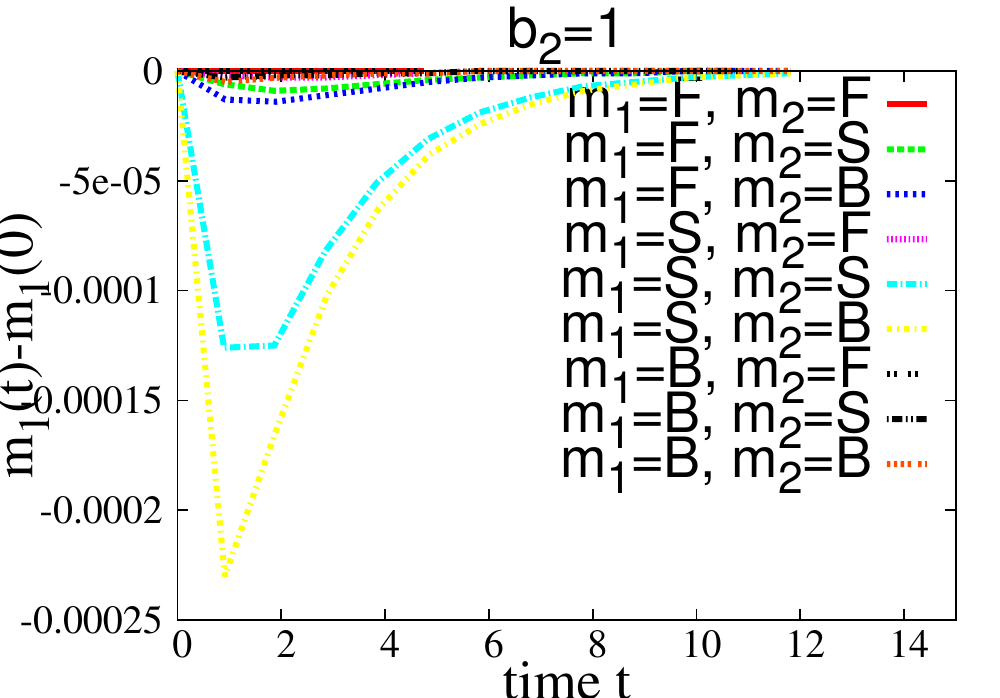}
%\caption{{\bf transcription perturbation:} time dependent response of cerna 1, with after the perturbation $\delta b_2=0.8$ at $t=0$($b_\mu=1=\mu^{SS}$) Cernas have $k^+=100,1,0.01$: they are in the bound, susc and free regime respectively ($\epsilon=0.01$)}
\includegraphics[width=0.23\textwidth]{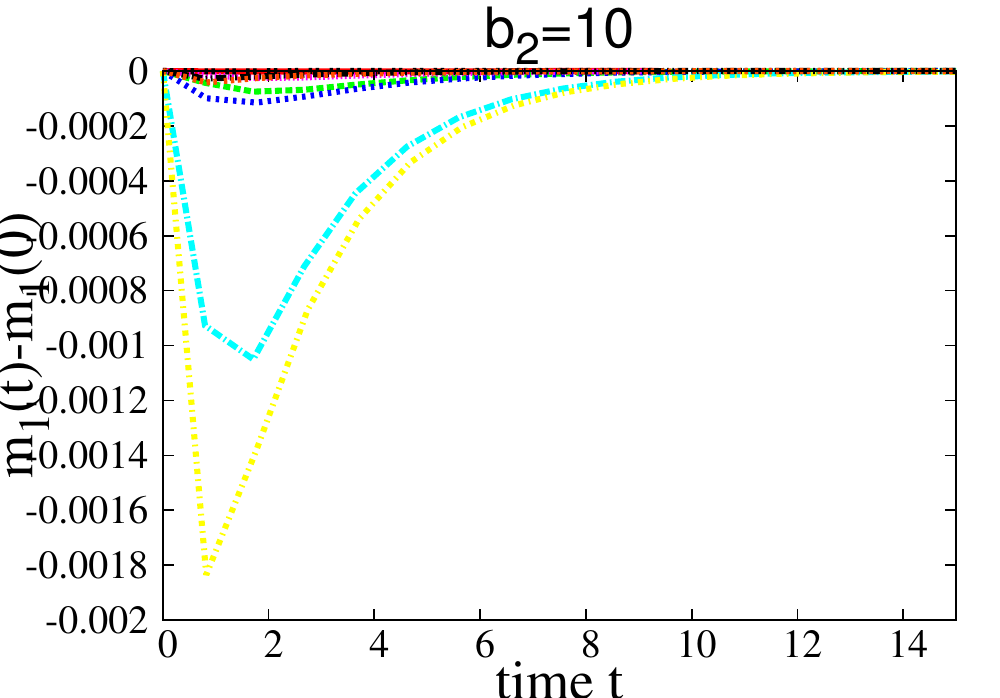}\\
%\caption{{\bf transcription perturbation:}time dependent response of cerna 1, with after the perturbation $\delta b_2=8.4$ at $t=0$($b_\mu=1=\mu^{SS}$)  }
\includegraphics[width=0.23\textwidth]{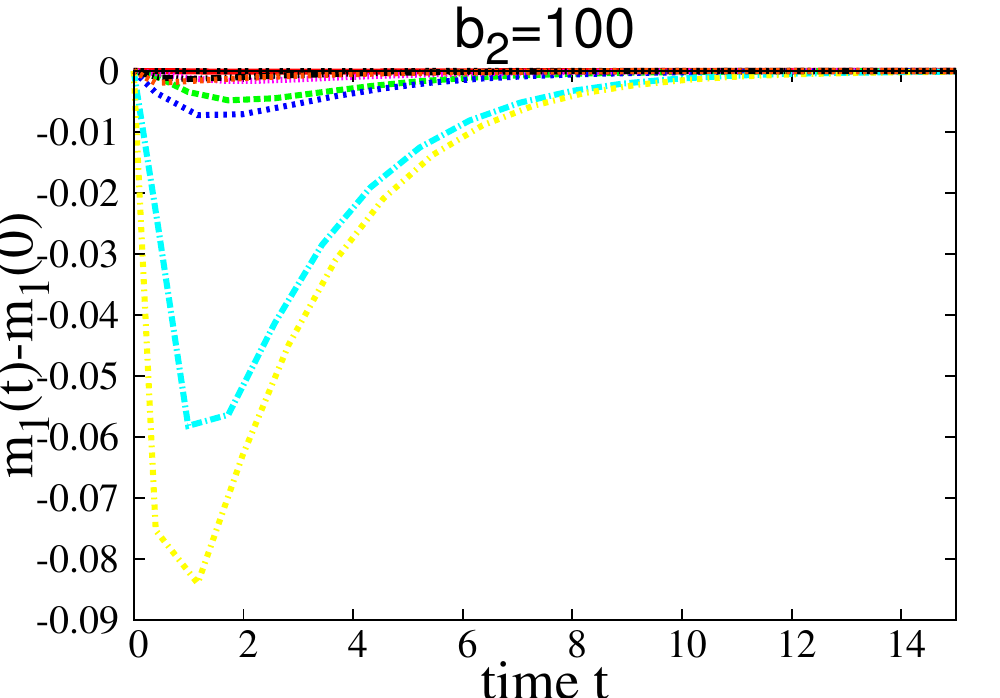}
%\caption{{\bf transcription perturbation:}time dependent response of cerna 1, with after the perturbation $\delta b_2=50$ at $t=0$($b_\mu=1=\mu^{SS}$)  }
\includegraphics[width=0.23\textwidth]{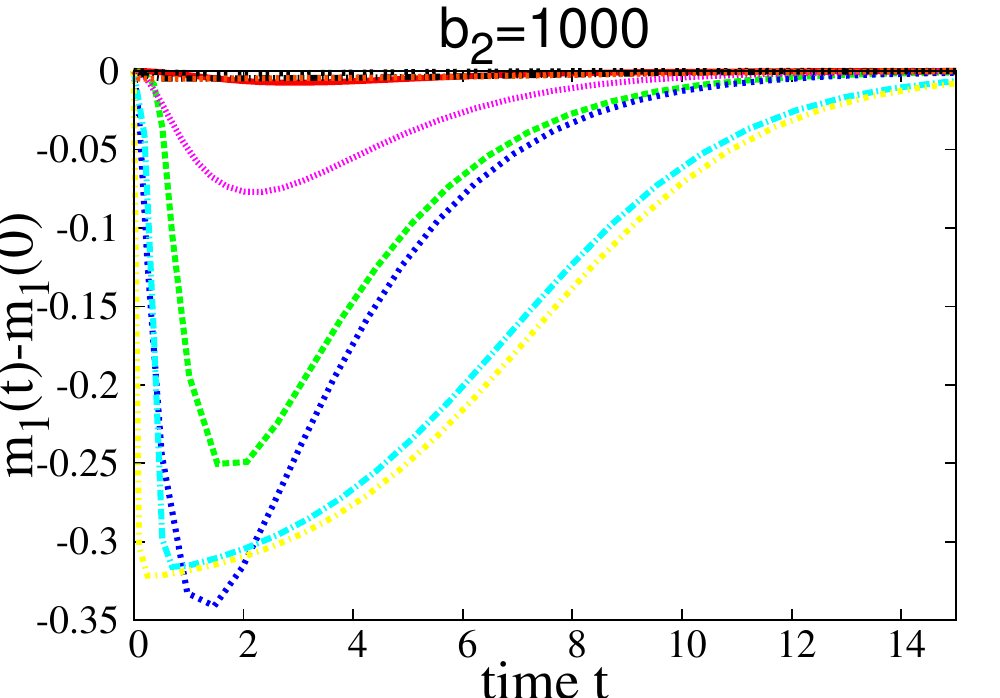}
%\caption{{\bf transcription perturbation:}time dependent response of cerna 1, with after the perturbation $\delta b_2=800$ at $t=0$($b_\mu=1=\mu^{SS}$)  }
\caption{Dynamical evolution of ceRNA 1  ($\Delta m_i(t)=m_1(t)-m_1(0)$), after a transcriptional perturbation $\Delta_2$ of ceRNA 2 at $t=0$. ceRNAs have different binding rates so that they can be in the free (F), susceptible (S) or bound (B) regime. Note that for small negative perturbations (top left plot) a significant response is achieved only for the S-S and the S-B pairs. Increasing the perturbation size, F-F cross-talk is activated, as are the S-F, F-S and F-B interactions.
}
\label{IRdown2}
\end{figure}

Deviations from linear response behavior occur also in the case of negative perturbations (reductions of the transcription rates), as shown in Figure \ref{IRdown}, where the transcription rate of ceRNA 2 is set to zero for $t>0$, i.e. $\Delta_2=-1$. If the transcription rate $b_2$ before the perturbation is sufficiently large, cross-talk between susceptible ceRNAs increases slowly while the response by free ceRNAs increases. Figure \ref{IRdown2} shows indeed that the levels of free species are sensibly depleted over a finite time window. 

\begin{figure}
\centering
{\includegraphics[width=0.48\textwidth]{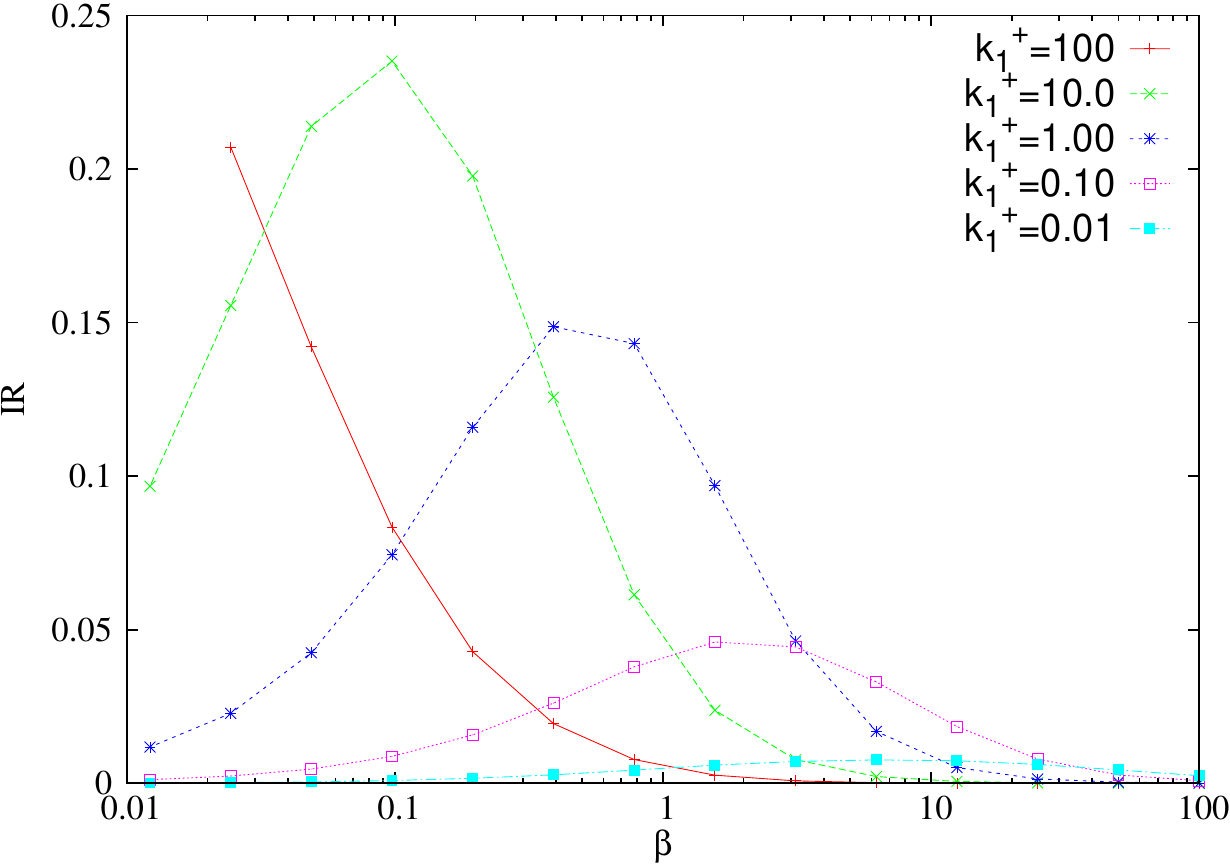}}
\caption{Integrated response IR of ceRNA 1 as a function of the miRNA transcription rate $\beta$. Remaining kinetic parameters are fixed as following: $b_1=\beta=1$, $b_2=1$, $d_1=d_2=\delta=1$, $k_1^-=k_2^-=0$, $\kappa_1=\kappa_2=1$, $k_2^+=1$, $\Delta=2$.
}
\label{selectivity}
\end{figure}

Quite remarkably, however  (see Figure \ref{selectivity}), selectivity is preserved also in case of large perturbations: cross-talk is activated only among a subset of ceRNAs, those whose binding kinetics lies in a finite window which depends on the perturbation size and on miRNA level. Others ceRNAs are almost unaffected by the perturbation.

\subsection{Threshold perturbation and saturation phenomena}

We have just seen that cross-talk between `bound' ceRNAs follows a threshold behaviour: if the perturbation is small, the response $\delta m_i$ grows linearly with the perturbation, according to $\delta m_i=\chi_{ij}\delta b_j$, where $\chi_{ij}$ is small as predicted by linear response theory; if however the perturbation overcomes a given threshold $\Delta_{th}$,  linear response theory break down and non-linear effects become important. Upon increasing further the perturbation, the IR returns to a linear behaviour, due to saturation effects. Intuitively, after a large positive perturbation that shifts the level of a ceRNA up, miRNAs are temporarily completely sequestered and other ceRNAs become completely free. Hence $\Delta m_i$  saturates to the maximal value, as shown in the bottom-right panel of figure \ref{IRup2}. Accordingly, the relaxation time (which depends weakly on the perturbation size in the linear response regime) increases linearly with the perturbation when $\Delta_j > \Delta_{th}$.

\begin{figure}
\centering
{\includegraphics[width=0.48\textwidth]{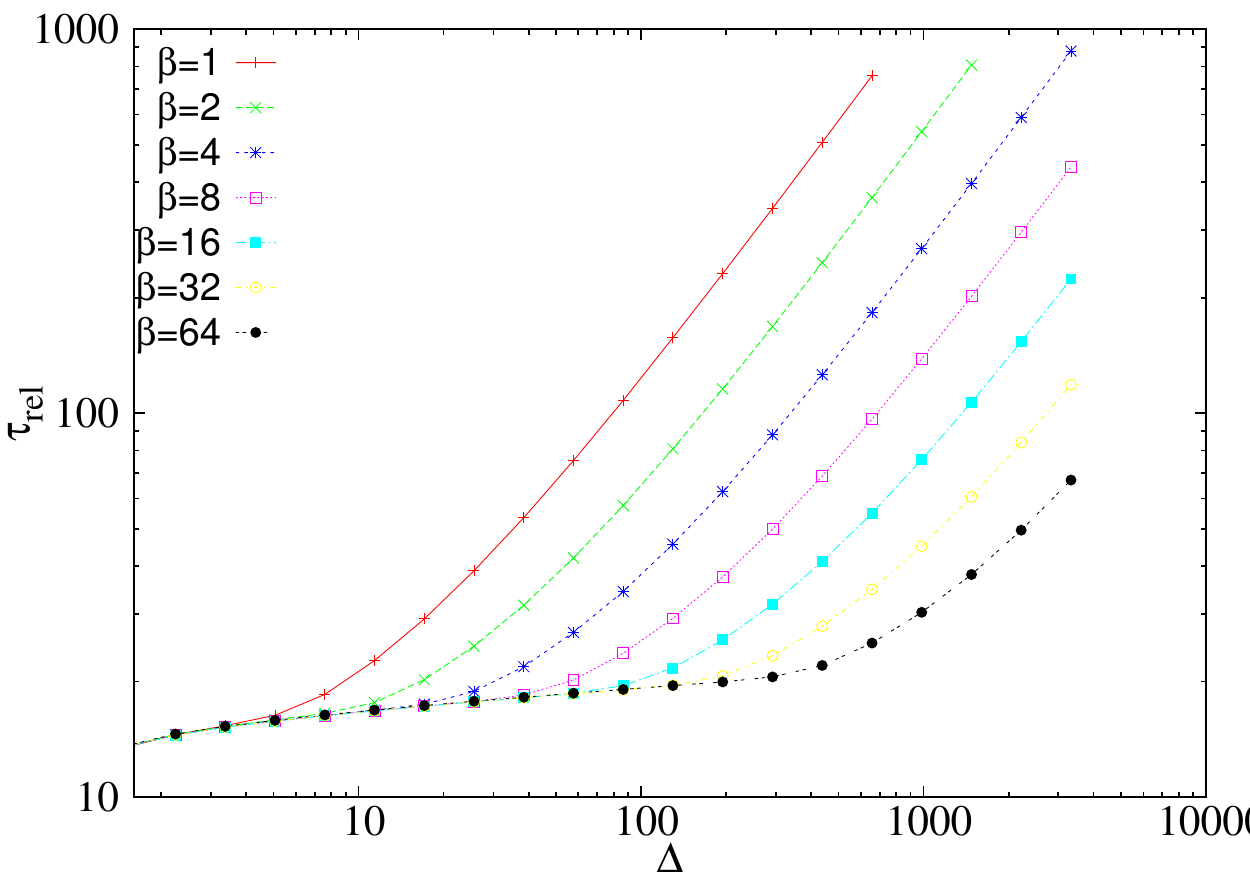}}
\caption{Relaxation time $\tau_{rel}$ as a function of the size of the perturbation, for different values of the rate of the miRNA transcription rate $\beta$. Remaining kinetic parameters are as follows: $b_1=\beta=1$, $b_2=1$, $d_1=d_2=\delta=1$,$k^+_1=k^+_2=100$, $k_1^-=k_2^-=0$.
}
\label{tau_beta}
\end{figure}

An estimate of the relaxation times after large, saturating perturbation can be worked out in the case of a kinetically homogeneous system (i.e. one in which binding kinetics is the same for all ceRNAs) assuming that ceRNAs and miRNAs are at equilibrium with respect to the instantaneous values of the levels of the complexes. One finds (see Supporting Text for details)
\begin{equation}
\tau_{rel} \simeq %\frac{k\sum_j m_j^\star}{2\beta \kappa}\mu_{0,i}\simeq\frac{\sum_j b_j+\Delta b}{2\beta \kappa}\simeq
\frac{ b_j \Delta_j}{2\beta \kappa}~~.
\end{equation}
Hence when $\Delta$ is very large relaxation times $\tau$ decreases upon increasing either the catalytic processing speed $\kappa$ or the miRNA transcription rate $\beta$. This is consistent with the numerical results shown in Figure \ref{tau_beta} and in Figure \ref{fig5s} of the Supporting Text. 

\begin{figure}
\centering
{\includegraphics[width=0.48\textwidth]{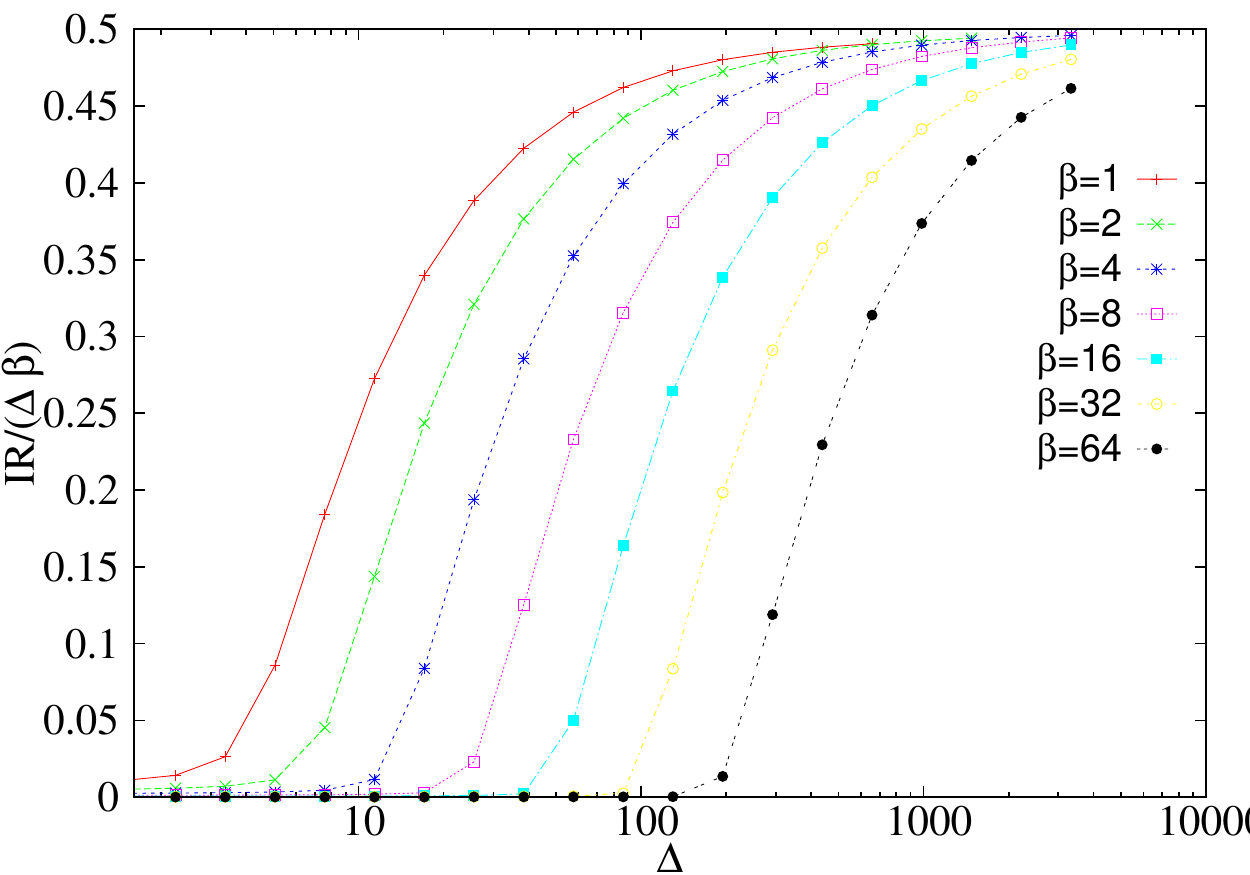}}
\caption{Integrated response IR between bound ceRNAs as a function of the perturbation size $\Delta$, for different miRNA transcription rates $\beta$. Remaining kinetic parameters are as follows: $b_1=b_2=1$, $b_2=1$, $d_1=d_2=\delta=1$, $k_1^-=k_2^-=0$, $\kappa_1=\kappa_2=1$, $k_2^+=100$
}
\label{threshold_beta}
\end{figure}

Notice also (see Figures \ref{threshold_beta} and Figure \ref{fig6s} of the Supporting Text) that not only relaxation times, but also the value of the threshold $\Delta_{th}$ appears to shift upon varying $\beta$ and $\kappa$.
%\red{according to computation $\Delta_{th}\sim [\mu]^2 $ why? stima in appendix}  \red{stima di $\tau$ in appendix?, saturation law in appendix?}

\subsection{Responsiveness and amplification}

Velocity in changing molecular levels in response to a perturbation (what we shall call `responsiveness' here) can be a desireable feature in cells: for instance, differentiation processes typically need rapid shifts in the levels of specific molecules. The ease of synthesis of small RNA molecules may be beneficial for quick response/adaptation to environmental stress \citep{altuvia2000switching}, and it has been quantitatively shown that PTR is advantageous precisely when fast responses to external signals are required \citep{shimoni2007regulation}. We are here in the position to compare the properties of transcriptional regulation by perturbation of transcription rate of a given gene to those of direct PTR by perturbation of miRNA transcription rates and indirect PTR by perturbation of the transcription rate of a competitor of the gene. Our goal is to quantify the differences between the three regulation modes, to pin down the situations when regulating through the ceRNA effect can be more effective.

\begin{figure}
\centering
{\includegraphics[width=0.48\textwidth]{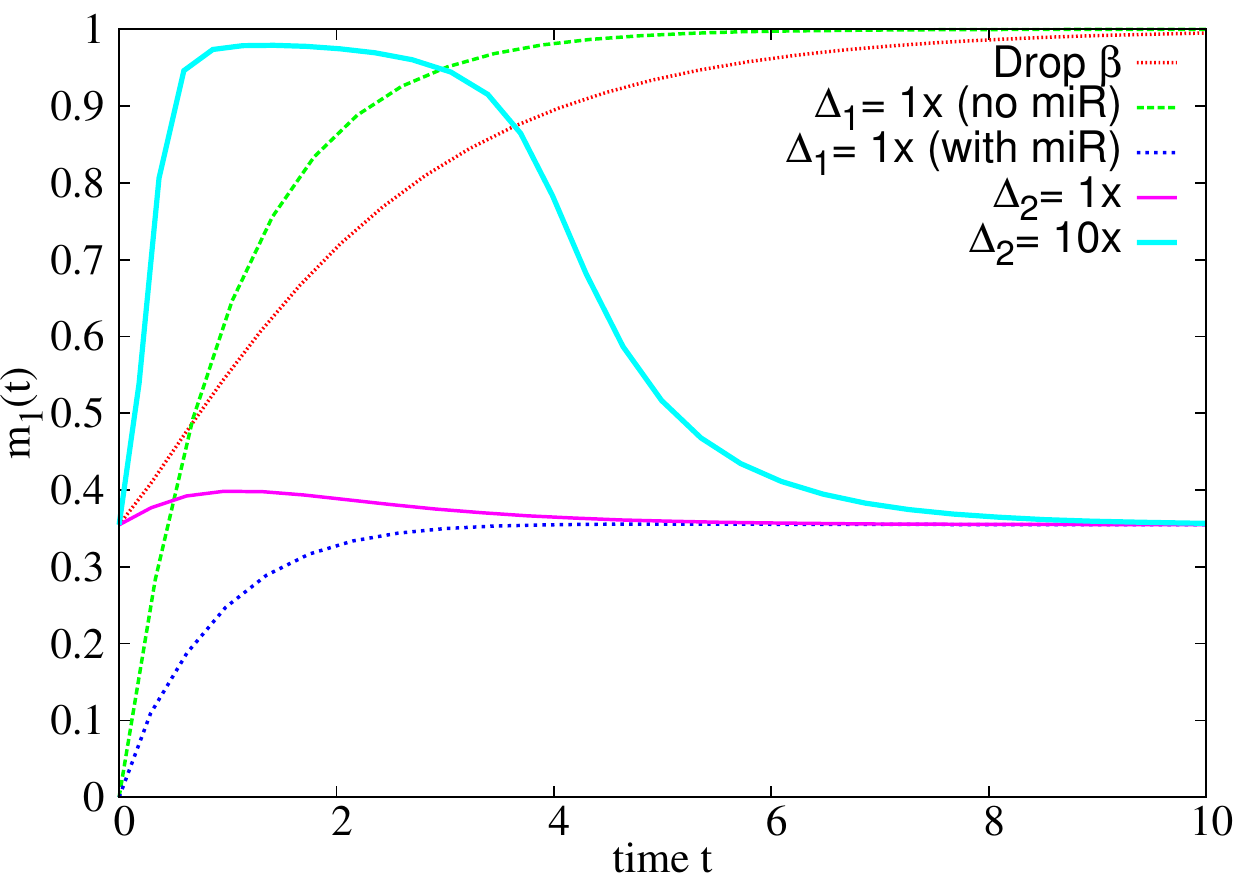}}
{\includegraphics[width=0.48\textwidth]{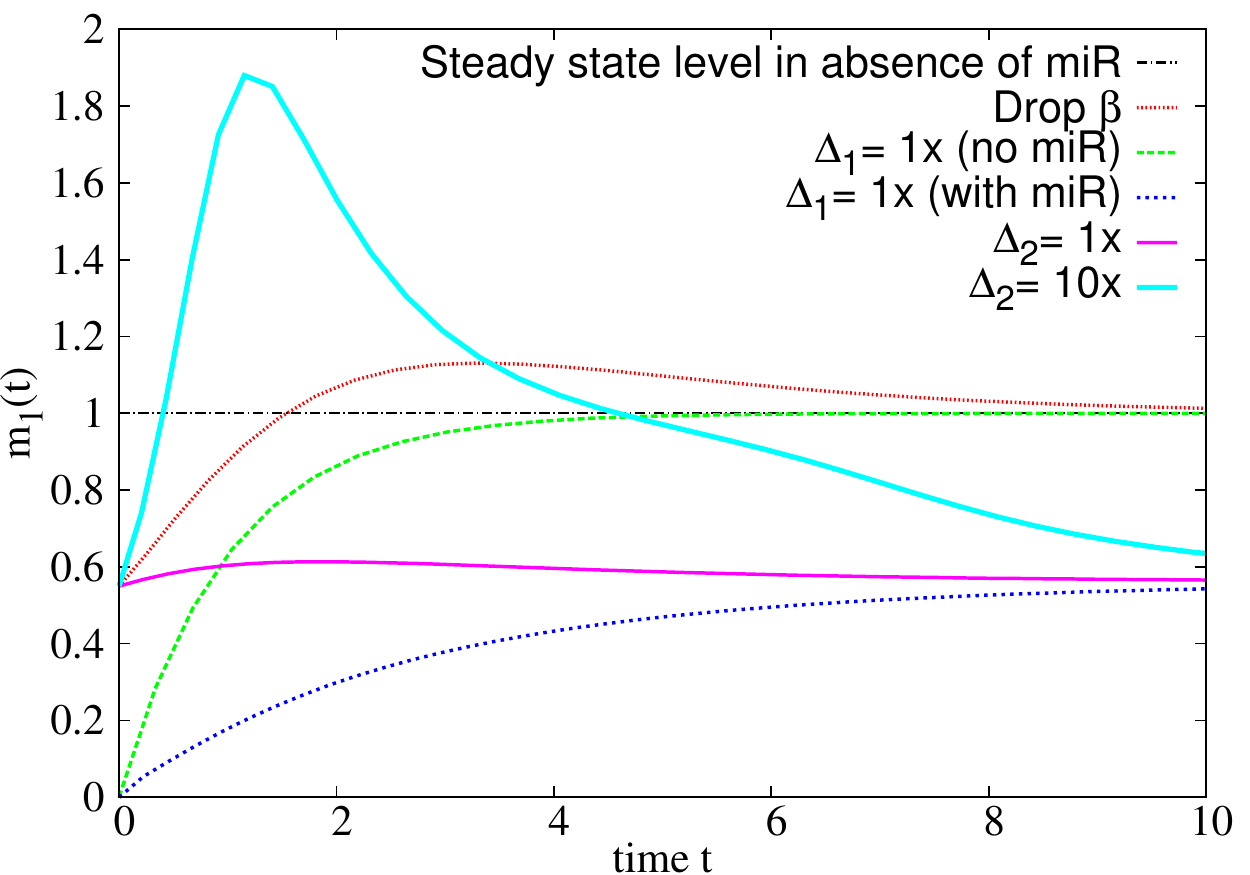}}
\caption{Time evolution of $m_1$ after different kinds of perturbations. The comparison is carried out in case of fast complex processing (top panel, $\kappa=1$) and slow complex processing (bottom panel, $\kappa=0.2$). Different perturbations are indicated as follows. \textit{Drop $\beta$}: miRNA transcription rate $\beta$ set to zero at $t>0$; \textit{$\Delta_1=1\times$ (no miR)}: activation of transctiption of the gene with $b_1=1$ at $t>0$ in absence of miR; \textit{$\Delta_1=1\times$ (with miR)}: activation of transctiption of the gene with $b_1=1$ at $t>0$  in presence of miRNA;  \textit{$\Delta_2=1\times$}: activation of transctiption of the ceRNA with $b_2=1$ at $t>0$; \textit{$\Delta_2=10\times$}: activation of transctiption of the ceRNA with $b_2=10$ at $t>0$. Remaining parameters are as follows: $k^-=0$ $b_1=1$, $\beta=2$, $d_1=d_2=\delta=1$, $k^-=10$, $k_1^+=10$, $k_2^+=1000$. \label{responsive_up}
}
\end{figure}

\begin{figure}
\centering
{\includegraphics[width=0.48\textwidth]{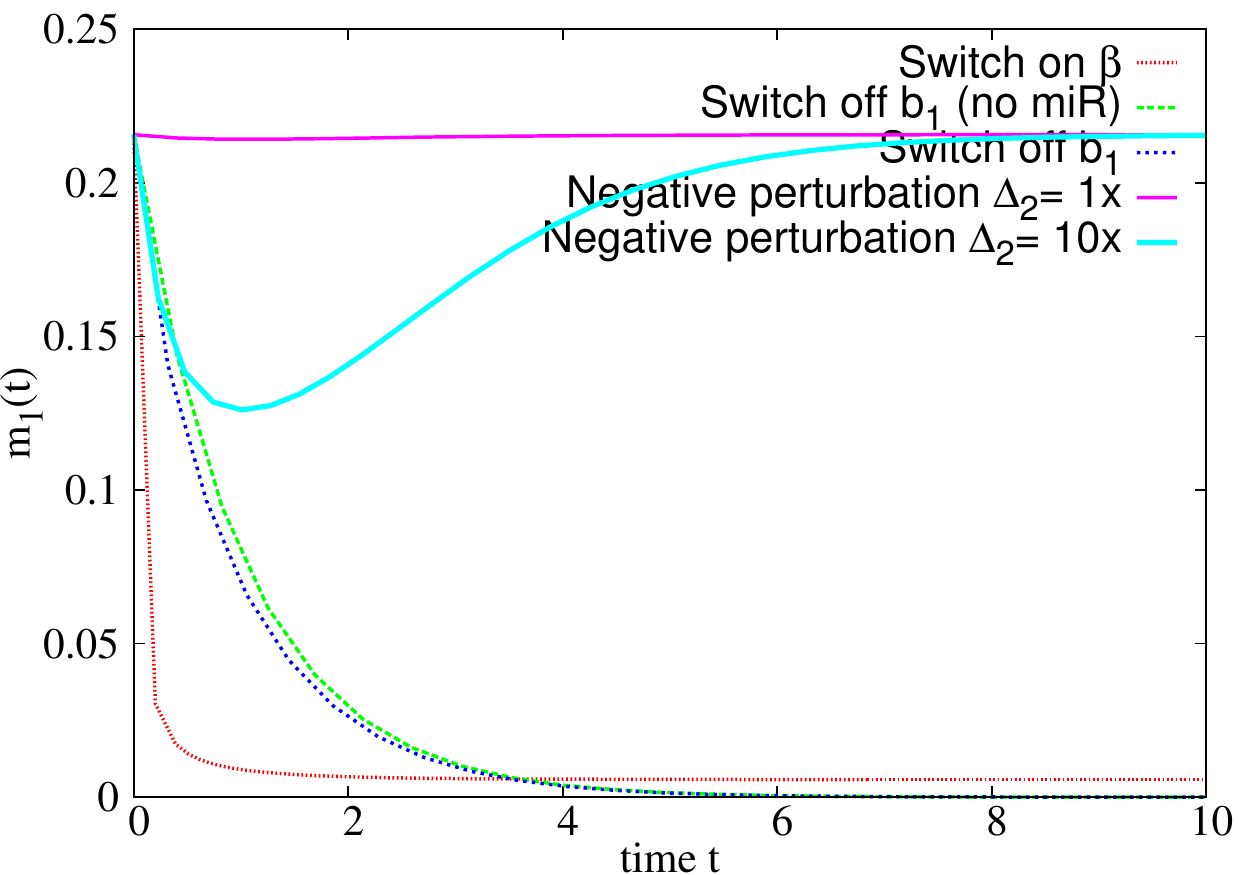}}
\caption{{\bf Gene silencing.} Time evolution of $m_1$ after different kinds of perturbations. Legends are as follows. \textit{Switch on $\beta$}: miRNA transcription rate $\beta$ activated at $t>0$; \textit{Switch off $b_1$ (no miR)}: block the transctiption of gene 1 at $t>0$ in absence of miRNA; \textit{Switch off $b_1$}: block the transctiption of gene 1 at $t>0$ in presence of miRNA; \textit{Negative perturbation $\Delta_2=1\times$}: block transctiption of the ceRNA 2 with $b_2=1$ at $t>0$; \textit{Negative perturbation $\Delta_2=10\times$}: block transctiption of ceRNA 2 with $b_2=10$ at $t>0$. Remaining parameters are as follows: $b_1=1$, $\beta=4$, $d_1=d_2=\delta=1$, $k^-=10$, $k_1^+=10$, $k_2^+=1000$, $\kappa=1$.
}
\label{responsive_down}
\end{figure}

Figure \ref{responsive_up} shows that switching off the transcription of a miRNA may not be the fastest way to increase the level of a transcript ($m_1$ in this case), because it takes some time for miRNA to be eliminated. On the other hand, turning on the transcription of the gene in absence of miRNAs or increasing the transcription of a ceRNA by several folds both result in rapid increase of the level of the gene.
Interestingly, if complexes are close to equilibrium and their lifetime is longer than that of free molecules, i.e. if the condition for cross-talk amplification are met, a sudden increase of the transcription rate of a ceRNA may temporarily bring the level of the gene above the steady state value in absence of miRNA  (see bottom panel in Figure \ref{responsive_up}). This effect is due to the massive release of free molecules from the dissociation of a large number of complexes $c_i$ right after the perturbation, and it is more pronounced if the affinity of the ceRNA is higher than that of the gene ($k_2^+>k_1^+$) as in the case considered in Figure \ref{responsive_up}. On the other hand, in Figure  \ref{responsive_down} it is shown that the fastest way to reduce the expression level of a gene is to turn on the transcription rate of the miRNA, while decreasing either the transcription rate of the gene or that of a competitor seem to imply a slower response.

\begin{comment}
\red{attempt of derivarion
If cernas are at equilibrium: $m=m(\mu)$,
hence 
\[
\dot{m_1}=\frac{\partial m_1}{\partial \mu}\dot{\mu}
\]
Susc cerna respond fast to large perturbations!
 if ceRNA2 is sequestring the mir $\dot{\mu}=-\Delta_2$???
di quello che succede modificandoi parametri
di perche ha senso che uncerna sia trascrito piu velocemente del gene
}
\end{comment}

\section{Discussion}

Considering the time scales involved, it is clear that dynamical effects may play an important role in PTR. Quantifying their relevance in comparison to steady state phenomenology may on one hand help to interpret experimental results, and on the other provide an overall understanding of the competition mechanism by which cells may achieve selective control of PTR through miRNAs. We have extended here the steady state analysis of the miRNA-ceRNA interaction network by studying the dynamics and response of a system of post-transcriptionally regulated RNAs, focusing on the linearized dynamics in the limit of small perturbations, and on numerical analysis in the case of large perturbations.
 
Our results can be summarized as follows. While steady state cross-talk scenario requires that miRNA-ceRNA complexes decay, at least partially, through a stoichiometric channel of degradation \citep{figliuzzi2013micrornas}, cross-talk can be effective even in complete absence of stoichiometric processing if the system is away from stationarity. Quite importantly, cross-talk can be dynamically amplified when the processing of miRNA-ceRNA complexes is slower than spontaneous ceRNA degradation: in this situation the dynamical response may even overcome the steady state response (on sufficiently short time scales) with fully-stoichiometric complex processing. Therefore, the emergent cross-talk scenario found in \citep{figliuzzi2013micrornas} at the steady state occurs, possibly enhanced, even in transients.

On the other hand, the response to large perturbations can be strongly non-linear, and a kind of `extended cross-talk' appears above a specific threshold perturbation: in this situation, non only can susceptible-susceptible and susceptible-bound ceRNA pairs interact, but also other pairs of ceRNA may effectively interact. Most notably, bound-bound and free-free ceRNA pairs may cross-talk in transients. When the perturbation is particularly large, the system saturates, as relaxation times increase linearly with the perturbation size. The size of the transcriptional perturbation ultimately determines the width of the time window for which cross-talk is active. 

Finally, we have shown that the ceRNA effect provides a mechanism by which a cell may achieve fast positive shifts in the level of a ceRNA when necessary; obtaining rapid negative shifts in the same way is instead less efficient, as the fastest decrease in RNA levels is obtained by increasing the level of the miRNA.

It is worth remarking that a major feature of the steady-state scenario, namely the emergence of selectivity, is fully preserved dynamically, so that target specificity in the ceRNA cross-talk network is ensured even away from the steady state. Likewise, by the ceRNA effect one may obtain different cross-talk networks upon changing the miRNA levels, suggesting that the so-called `miR programs' may be a viable and effective mechanism to regulate the transcriptome composition on physiological time scales.

The small-scale model discussed here (two ceRNAs, one miRNA) gives many clues about the regulatory potential of the ceRNA effect. Still, it would be important to explore this scenario on a large-scale miRNA-ceRNA networks, where topological as well as kinetic ingredients may provide further insight on why PTR by small RNAs is so ubiquitous. In addition, it should always be kept in mind that signalling in this context can be limited by noise \citep{detwiler2000engineering,mehta2008quantitative}. It has been shown that the noise profiles of microRNA-regulated genes are almost identical in the case of stoichiometric and catalytic complex processing \citep{Noorbakhsh2013intrinsic,bosia2013modelling}. This suggests that dynamical considerations are crucial to further understand how noise processing may be performed during PTR. Further work on the emergence and properties of ceRNA cross-talk networks may therefore prove to yield deeper insights on a number of key issues for transcriptome and proteome regulation.

\bibliographystyle{unsrt}

\bibliography{bib_dynamics}

\begin{thebibliography}{10}

\bibitem{chekulaeva2009mechanisms}
M.~Chekulaeva and W.~Filipowicz.
\newblock Mechanisms of mi{RNA}-mediated post-transcriptional regulation in
  animal cells.
\newblock {\em Curr. Opin. Cell Biol.}, 21:452--460, 2009.

\bibitem{valencia2006control}
M.~A. Valencia-Sanchez, J.~Liu, G.~J. Hannon, and R.~Parker.
\newblock Control of translation and m{RNA} degradation by mi{RNA}s and
  si{RNA}s.
\newblock {\em Genes Dev.}, 20:515--524, 2006.

\bibitem{bartel2004micrornas}
D.~P. Bartel.
\newblock Micro{RNA}s: genomics, biogenesis, mechanism, and function.
\newblock {\em Cell}, 116:281--297, 2004.

\bibitem{tsang2007microrna}
J.~Tsang, J.~Zhu, and A.~van Oudenaarden.
\newblock Micro{RNA}-mediated feedback and feedforward loops are recurrent
  network motifs in mammals.
\newblock {\em Mol. Cell}, 26:753--767, 2007.

\bibitem{osella2011role}
M.~Osella, C.~Bosia, D.~Cor{\'a}, and M.~Caselle.
\newblock The role of incoherent micro{RNA}-mediated feedforward loops in noise
  buffering.
\newblock {\em PLoS Comp. Biol.}, 7:e1001101, 2011.

\bibitem{arvey2010target}
A.~Arvey, E.~Larsson, C.~Sander, C.~S. Leslie, and D.~S. Marks.
\newblock Target m{RNA} abundance dilutes micro{RNA} and si{RNA} activity.
\newblock {\em Mol. Sys. Biol.}, 6:363, 2010.

\bibitem{salmena2011cerna}
L.~Salmena, L.~Poliseno, Y.~Tay, L.~Kats, and P.~P. Pandolfi.
\newblock A ce{RNA} hypothesis: the rosetta stone of a hidden {RNA} language?
\newblock {\em Cell}, 146:353--358, 2011.

\bibitem{cesana2011long}
M.~Cesana, D.~Cacchiarelli, I.~Legnini, T.~Santini, O.~Sthandier, M.~Chinappi,
  A.~Tramontano, and I.~Bozzoni.
\newblock A long noncoding {RNA} controls muscle differentiation by functioning
  as a competing endogenous {RNA}.
\newblock {\em Cell}, 147:358--369, 2011.

\bibitem{karreth2011vivo}
F.~A. Karreth, Y.~Tay, D.~Perna, U.~Ala, S.~M. Tan, A.~G. Rust, G.~DeNicola,
  K.~A. Webster, D.~Weiss, P.~A. Perez-Mancera, M.~Krauthammer, R.~Halaban,
  P.~Provero, D.~J. Adams, D.~A. Tuveson, and P.~P. Pandolfi.
\newblock In vivo identification of tumor-suppressive {PTEN} ce{RNA}s in an
  oncogenic {BRAF}-induced mouse model of melanoma.
\newblock {\em Cell}, 147:382--395, 2011.

\bibitem{tay2011coding}
Y.~Tay, L.~Kats, L.~Salmena, D.~Weiss, S.~M. Tan, U.~Ala, F.~Karreth,
  L.~Poliseno, P.~Provero, F.~Di~Cunto, J.~Lieberman, I.~Rigoutsos, and P.~P.
  Pandolfi.
\newblock Coding-independent regulation of the tumor suppressor {PTEN} by
  competing endogenous m{RNA}s.
\newblock {\em Cell}, 147:344--357, 2011.

\bibitem{figliuzzi2013micrornas}
M.~Figliuzzi, E.~Marinari, and A.~De~Martino.
\newblock Micro{RNA}s as a selective channel of communication between competing
  rnas: a steady-state theory.
\newblock {\em Biophys. J.}, 104:1203--1213, 2013.

\bibitem{ala2013integrated}
U.~Ala, F.~A. Karreth, C.~Bosia, A.~Pagnani, R.~Taulli, V.~L{\'e}opold, Y.~Tay,
  P.~Provero, R.~Zecchina, and P.~P. Pandolfi.
\newblock Integrated transcriptional and competitive endogenous {RNA} networks
  are cross-regulated in permissive molecular environments.
\newblock {\em Proc. Nat. Acad. Sci. USA}, 110:7154--7159, 2013.

\bibitem{argaman2000fhla}
L.~Argaman and S.~Altuvia.
\newblock fhl{A} repression by {O}xy{S RNA}: kissing complex formation at two
  sites results in a stable antisense-target {RNA} complex.
\newblock {\em J. Mol. Biol.}, 300:1101--1112, 2000.

\bibitem{wagner200212}
E.~G. Wagner, S.~Altuvia, and P.~Romby.
\newblock Antisense rnas in bacteria and their genetic elements.
\newblock {\em Adv. Genetics}, 46:361--398, 2002.

\bibitem{haley2004kinetic}
B.~Haley and P.~D. Zamore.
\newblock Kinetic analysis of the {RNA}i enzyme complex.
\newblock {\em Nature Struct. Mol. Biol.}, 11:599--606, 2004.

\bibitem{alon2007introduction}
U.~Alon.
\newblock {\em Introduction to Systems Biology: Design Principles of Biological
  Networks}.
\newblock CRC press, 2007.

\bibitem{wang2010toward}
X.~Wang, Y.~Li, X.~Xu, and Y.-H. Wang.
\newblock Toward a system-level understanding of micro{RNA} pathway via
  mathematical modeling.
\newblock {\em Biosystems}, 100:31--38, 2010.

\bibitem{morris2004small}
K.~V. Morris, S.~Chan, S.~E Jacobsen, and D.~J. Looney.
\newblock Small interfering {RNA}-induced transcriptional gene silencing in
  human cells.
\newblock {\em Science}, 305:1289--1292, 2004.

\bibitem{altuvia2000switching}
S.~Altuvia and E.~G. Wagner.
\newblock Switching on and off with {RNA}.
\newblock {\em Proc. Nat. Acad. Sci. USA}, 97:9824--9826, 2000.

\bibitem{shimoni2007regulation}
Y.~Shimoni, G.~Friedlander, G.~Hetzroni, G.~Niv, S.~Altuvia, O.~Biham, and
  H.~Margalit.
\newblock Regulation of gene expression by small non-coding {RNA}s: a
  quantitative view.
\newblock {\em Mol. Sys. Biol.}, 3:138, 2007.

\bibitem{detwiler2000engineering}
P.~B. Detwiler, S.~Ramanathan, A.~Sengupta, and B.~I. Shraiman.
\newblock Engineering aspects of enzymatic signal transduction: photoreceptors
  in the retina.
\newblock {\em Biophys. J.}, 79:2801--2817, 2000.

\bibitem{mehta2008quantitative}
P.~Mehta, S.~Goyal, and N.~S. Wingreen.
\newblock A quantitative comparison of s{RNA}-based and protein-based gene
  regulation.
\newblock {\em Mol. Sys. Biol.}, 4:221, 2008.

\bibitem{Noorbakhsh2013intrinsic}
J.~Noorbakhsh, A.~H. Lang, and P.~Mehta.
\newblock Intrinsic noise of micro{RNA}-regulated genes and the ce{RNA}
  hypothesis.
\newblock {\em PLoS ONE}, 8:e72676, 2013.

\bibitem{bosia2013modelling}
C.~Bosia, A.~Pagnani, and R.~Zecchina.
\newblock Modelling competing endogenous {RNA} networks.
\newblock {\em PLoS ONE}, 8:e66609, 2013.

\end{thebibliography}

\newpage

\section*{Supporting Text}

\subsection*{Analysis of the linearized dynamics}

In Fourier space (where $\hat{a}(\omega)$ denotes the Fourier transform of $a(t)$) the  dynamics defined in Eq. (5) of the main text takes the form
%\begin{gather}
%i\omega\hat{x_i}=-d_i\hat{x_i} + \hat{\Delta b_i}-k_i^+([\mu]\hat{x_i}+[m_i]\hat{y})+k_i^-\hat{z_i}\nonumber\\
%i\omega\hat{y}=-\delta\hat{y} + \hat{\Delta\beta}-\sum_i k_i^+([\mu]\hat{x_i}+[m_i]\hat{y})+\sum_i(k_i^-+\kappa_i)\hat{z_i} \label{fouriereqs} \\
%i\omega\hat{z_i}=-(\sigma_i+k_i^-+\kappa_i)\hat{z_i} + k_i^+([\mu]\hat{x_i}+[m_i]\hat{y})\nonumber~~.
%\end{gather}
%or, in turn
\begin{gather}
\hat{x_i}(\omega)=\frac{\hat{\btil_i}-k_i^+\Gamma_i(\omega)[m_i]\hat{y}}{\ii \omega+d_i+k_i^+[\mu]\Gamma_i(\omega)}\nonumber\\
\hat{y}(\omega)=\frac{\hat{\betil}-[\mu]\sum_i k_i^+\Lambda_i(\omega)\hat{\btil_i}}{\Delta(\omega)}\\
\hat{z_i}(\omega)=\frac{k_i^+([\mu]\hat{x_i}+[m_i]\hat{y})}{\ii \omega+\kappa_i+\sigma_i+k_i^-}\nonumber~~,
\end{gather}
where
\begin{gather}
\Gamma_i(\omega)%=\frac{k_i^-+\sigma_i+\kappa_i}{\sigma_i+\kappa_i}\frac{1+i\omega\tau_{3,i}}{1+i\omega\tau_{2,i}}
=(1+\phi_i)\frac{1+\ii\omega\tau_{3,i}}{1+\ii\omega\tau_{2,i}} \nonumber\\
\Delta(\omega)=%i\omega+\delta+\sum_ik_i^+[m_i]\frac{i\omega+d_i}{i\omega+d_i+k_i^+[\mu]\Gamma_i(\omega)}=\\
\ii\omega+\delta+\sum_ik_i^+[m_i]\left(1+\frac{[\mu]}{\mu_{0,i}Z_i(\omega)}\right)^{-1}
\end{gather}
\begin{multline}
\Lambda_i(\omega)=%\frac{i\omega +\sigma_i}{(i\omega+d_i+k_i^+[\mu]\Gamma_i(\omega))(i\omega+\kappa_i+\sigma_i+k_i^-)}=\\
\frac{\sigma_i}{d_i(k_i^-+\kappa_i+\sigma_i)}\frac{1+\ii\omega \tau_{4,i}}{(1+\ii\omega\tau_{1,i})(1+\ii\omega\tau_{2,i})}\left(1+\frac{[\mu]}{\mu_{0,i}Z_i(\omega)}\right)^{-1} ~~,\nonumber 
\end{multline}
and we have used the time scales $\tau_{k,i}$ ($k=1,\ldots, 4$) as well as the function
\begin{equation}
Z_i(\omega)=\frac{(1+\ii\omega\tau_{1,i})(1+\ii\omega\tau_{2,i})}{1+\ii\omega\tau_{3,i}}~~,
\end{equation}
and the parameter
\begin{equation}
%F_i([\mu])=\frac{\mu_{0,i}}{[\mu]+\mu_{0,i}}~~~~~,~~~~~
\phi_i=\frac{k_i^-}{\sigma_i+\kappa_i}.%~~~~~,~~~~~\mu_{0,i}=\frac{d_i}{k_i^+}(1+\phi_i)
\end{equation}

The dynamical response may be quantified through the susceptibility
\begin{equation}
\hat{\chi_{ij}}(\omega)=\frac{\partial \hat{x_i}}{\partial \hat{\btil_j}}~~.
\end{equation}
We note that
\begin{equation}
\hat{\chi_{ij}}(\omega)=\frac{\partial \hat{x_i}}{\partial \hat{y}}\frac{\partial \hat{y}}{\partial \hat{\btil_j}}\equiv g_{i\mu}(\omega)g_{\mu j}(\omega)~~,
\end{equation}

where we have introduced the frequency-dependent gains 
%$g_{ij}(\omega)=g_{i\mu}(\omega)g_{\mu j}(\omega)$:
\begin{gather}
g_{i\mu}(\omega)\equiv \frac{\partial \hat{x_i}}{\partial \hat{y}}=%-\frac{k_i^+\Gamma_i(\omega)[m_i]}{i\omega+d_i+k_i^+[\mu]\Gamma_i(\omega)}=
-\frac{[m_i]}{[\mu]}\left(1+\frac{\mu_{0,i}}{[\mu]}Z_i(\omega)\right)^{-1}\\
g_{\mu j}(\omega)\equiv \frac{\partial \hat{y}}{\partial \hat{\btil_j}}=%-\frac{[\mu]k_j^+\Lambda_j(\omega)}{\Delta(\omega)}=
-\chi_{\mu\mu}(\omega)V_j(\omega)
\end{gather}
with $\chi_{\mu\mu}(\omega)=\Delta(\omega)^{-1}$ and
\begin{equation}
V_j(\omega)=
\begin{cases}
\frac{\sigma_j}{\sigma_j+\kappa_j}\frac{1+i\omega\tau_{4,j}}{1+i\omega\tau_{3,j}}\left(1+\frac{\mu_{0,j}}{[\mu]}Z_j(\omega)\right)^{-1}& \text{if $\sigma_j> 0$}\\
\frac{i\omega\tau_{5,j}}{1+i\omega\tau_{5,j}}\left(1+\frac{\mu_{0,j}}{[\mu]}Z_j(\omega)\right)^{-1}& \text{if $\sigma_j= 0$}~~.
\end{cases}
\end{equation}

~

Upon defining the filters $J_i(\omega), C_i(\omega), S_i(\omega)$ and $D(\omega)$ as in the Main Text, one may re-cast the above gains as
\begin{equation}
g_{i\mu}(\omega)=J_{i}(\omega)g_{i\mu}(0)
\end{equation}
and
\begin{equation}
g_{\mu j}(\omega)=
\begin{cases}
-D(\omega)J_j(\omega)S_j(\omega)g_{\mu j}(0) & \text{if $\sigma_j> 0$}\\
-D(\omega)J_j(\omega)C_j(\omega)\widetilde{g}_{\mu j}(0) & \text{if $\sigma_j= 0$}~~,
\end{cases}
\end{equation}
where $\widetilde{g}_{\mu j}(0)$ is the steady state term for the completely stoichiometric case (obtained upon setting $\kappa_i \rightarrow 0$ and $\sigma_i\rightarrow \kappa_i$)

Putting pieces together, we find
\begin{equation}\label{susk}
\hat{\chi_{ij}}(\omega)=
\begin{cases}
D(\omega)\Big[S_j(\omega)J_i(\omega)J_j(\omega)\Big]\chi_{ij}^{ss}%=\Psi_{ij}(\omega)\chi_{ij}^{SS}
& \text{if $\sigma_j\neq 0$}\\
D(\omega)\Big[C_j(\omega)J_i(\omega)J_j(\omega)\Big]\widetilde{\chi}_{ij}^{ss}%=\Phi_{ij}(\omega)\widetilde{\chi}_{ij}^{SS}
 &\text{if $\sigma_j= 0$}\\
\end{cases}
\end{equation}
where
\begin{equation}\label{sss}
\chi_{ij}^{ss}\equiv\hat{\chi_{ij}}(0)=g_{i\mu}(0)g_{\mu i}(0)
\end{equation}
and
\begin{equation}
\widetilde{\chi}_{ij}^{ss}\equiv\lim_{\sigma_j \to 0}\frac{\sigma_j+\kappa_j}{\sigma_j}\chi_{ij}^{ss}~~,%(\sigma_j)
\end{equation}
which corresponds to the steady state susceptibility of a system without recycling (i.e. with $\kappa_i \rightarrow 0$ and $\sigma_i\rightarrow \kappa_i$). Finally, the self response is given by
\begin{equation}
\hat{\chi_{ii}}(\omega)\equiv\frac{\partial \hat{x_i}}{\partial \hat{\btil_i}}=%\frac{1}{d_i}\frac{1}{1+i\omega\tau_{1,i}}\frac{1}{1+\frac{[\mu]}{\mu_{0,i}Z_i(\omega)}}=
\frac{J_i(\omega)\chi_{ii}(0)}{1+\ii\omega\tau_{1,i}}
\end{equation}

\subsection*{Susceptibilities}

\begin{figure}
\centering
%{\includegraphics[width=0.48\textwidth]{slow_phi.pdf}}\\
{\includegraphics[width=0.48\textwidth]{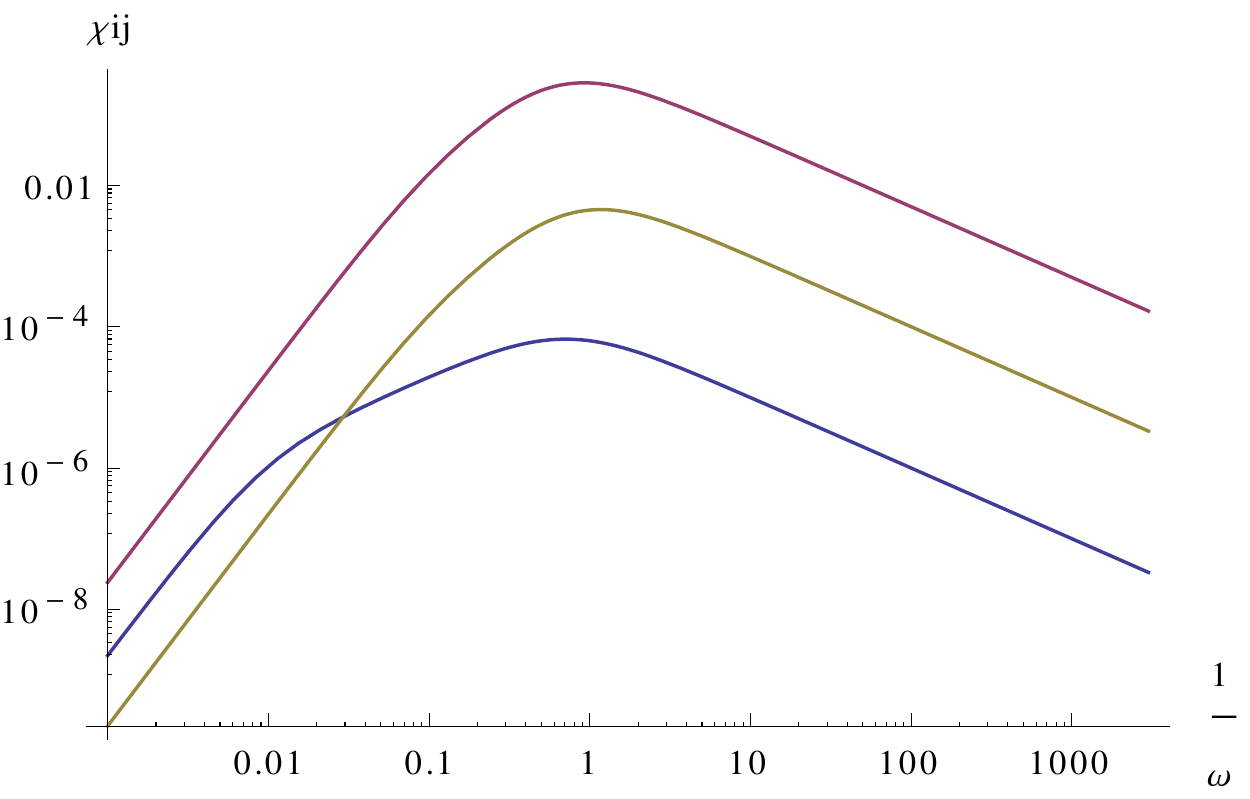}}
\caption{ {\bf Slow dissociation, fast processing} Dynamical susceptibility $\chi_{ij}(\omega)$ for slow complex dissociation in a fully catalitic system ($\sigma_i=0,\kappa_i=10$) for pairs of `free' ($\rho_i=100$, in yellow), `susceptible' ($\rho_i=1$, in red) and `bound' ($\rho_i=0.01$, in blue) ceRNAs. Remaining parameters are set as follows: $d_i=1, k_i^-=0, \delta=1, Z_i\equiv(1+\ii\omega\tau_{1,i})(1+\ii\omega\tau_{2,i})/(1+\ii\omega\tau_{3,i})=10$ for each $i$.\label{fig1s}}
\end{figure}

\begin{figure}
\centering
%{\includegraphics[width=0.48\textwidth]{recover_phi.pdf}}\\
{\includegraphics[width=0.48\textwidth]{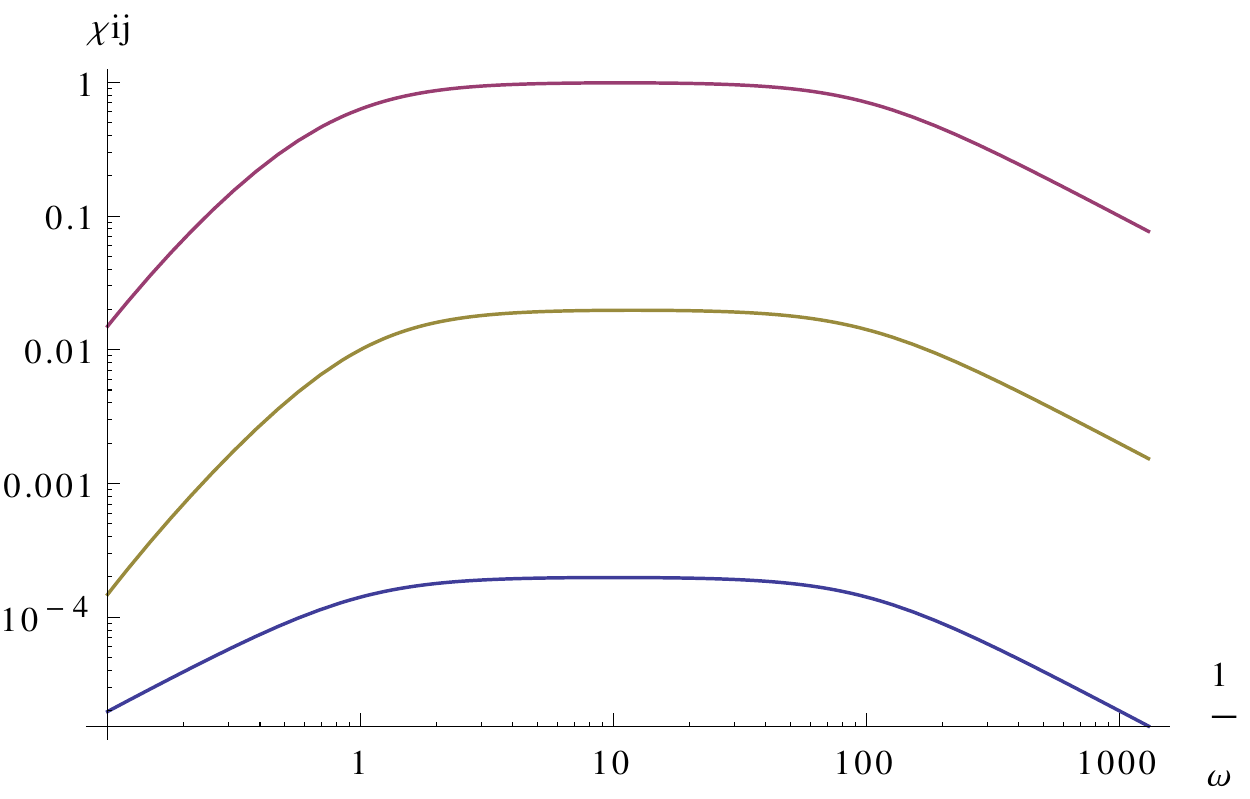}}
\caption{ {\bf Slow dissociation, slow processing} Dynamical susceptibility $\chi_{ij}(\omega)$ in a fully catalitic system ($\sigma_i=0,\kappa_i=0.01$) for a couple of free ceRNA ($\rho_i=100$, in yellow), for a couple of susc ceRNA ($\rho_i=1$, in red), for a couple of bound ceRNA ($\rho_i=0.01$, in blue). Other parameters are set as follows: $d_i=1, k_i^-=0, \delta=1, Z_i=10$ for each $i$.\label{fig2s}}
\end{figure}

\begin{figure}
\centering
%{\includegraphics[width=0.48\textwidth]{fast_phi.pdf}}\\
{\includegraphics[width=0.48\textwidth]{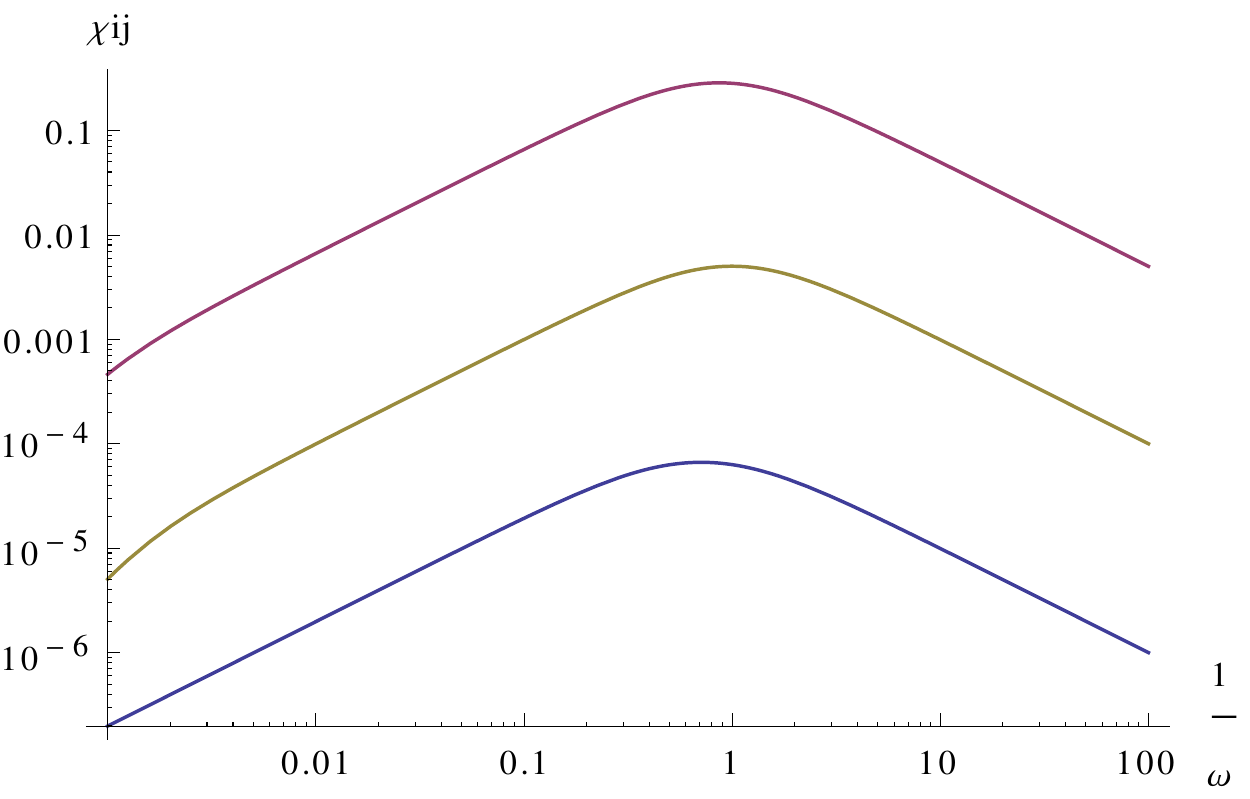}}
\caption{ {\bf Fast dissociation, fast processing} Dynamical susceptibility $\chi_{ij}(\omega)$ for fast complex dissociation in a fully catalitic system ($\sigma_i=0,\kappa_i=10$) for pairs of `free' ($\rho_i=100$, in yellow), `susceptible' ($\rho_i=1$, in red) and `bound' ceRNAs ($\rho_i=0.01$, in blue).  Other parameters are set as follows: $d_i=1, k_i^-=1000, \delta=1, Z_i=10$ for each $i$.\label{fig3s}}
\end{figure}

\begin{figure}
\centering
%{\includegraphics[width=0.48\textwidth]{amp_psi.pdf}}\\
{\includegraphics[width=0.48\textwidth]{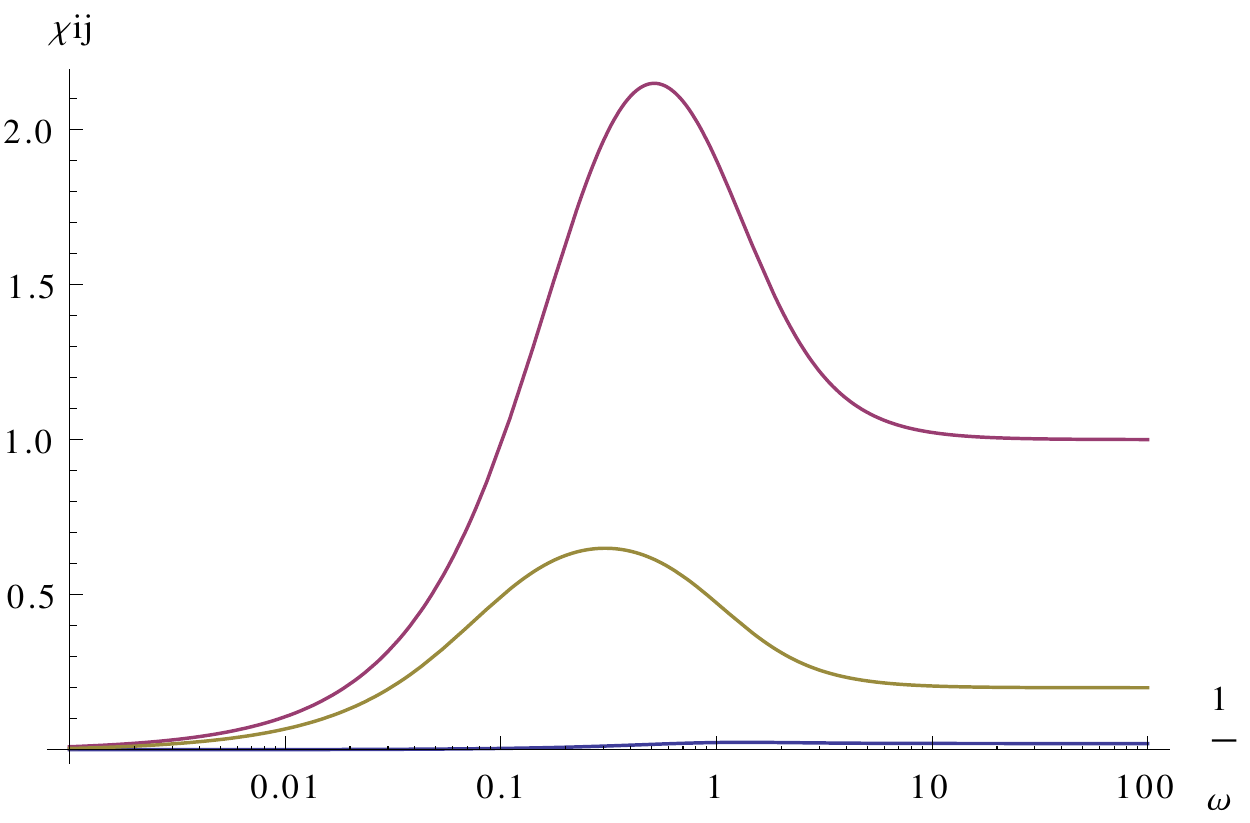}}
\caption{ {\bf Fast dissociation, slow processing} Dynamical susceptibility $\chi_{ij}(\omega)$ for fast complex dissociation in a fully stoichiometric system ($\sigma_i=0.5,\kappa_i=0$) for pairs of `free' ($\rho_i=100$, in yellow), `susceptible' ($\rho_i=1$, in red) and `bound' ($\rho_i=0.01$, in blue) ceRNAs. Other parameters are set as follows: $d_i=1, k_i^-=1000, \delta=1, Z_i=10$ for each $i$.\label{fig4s}}
\end{figure}

\begin{figure}
\centering
{\includegraphics[width=0.48\textwidth]{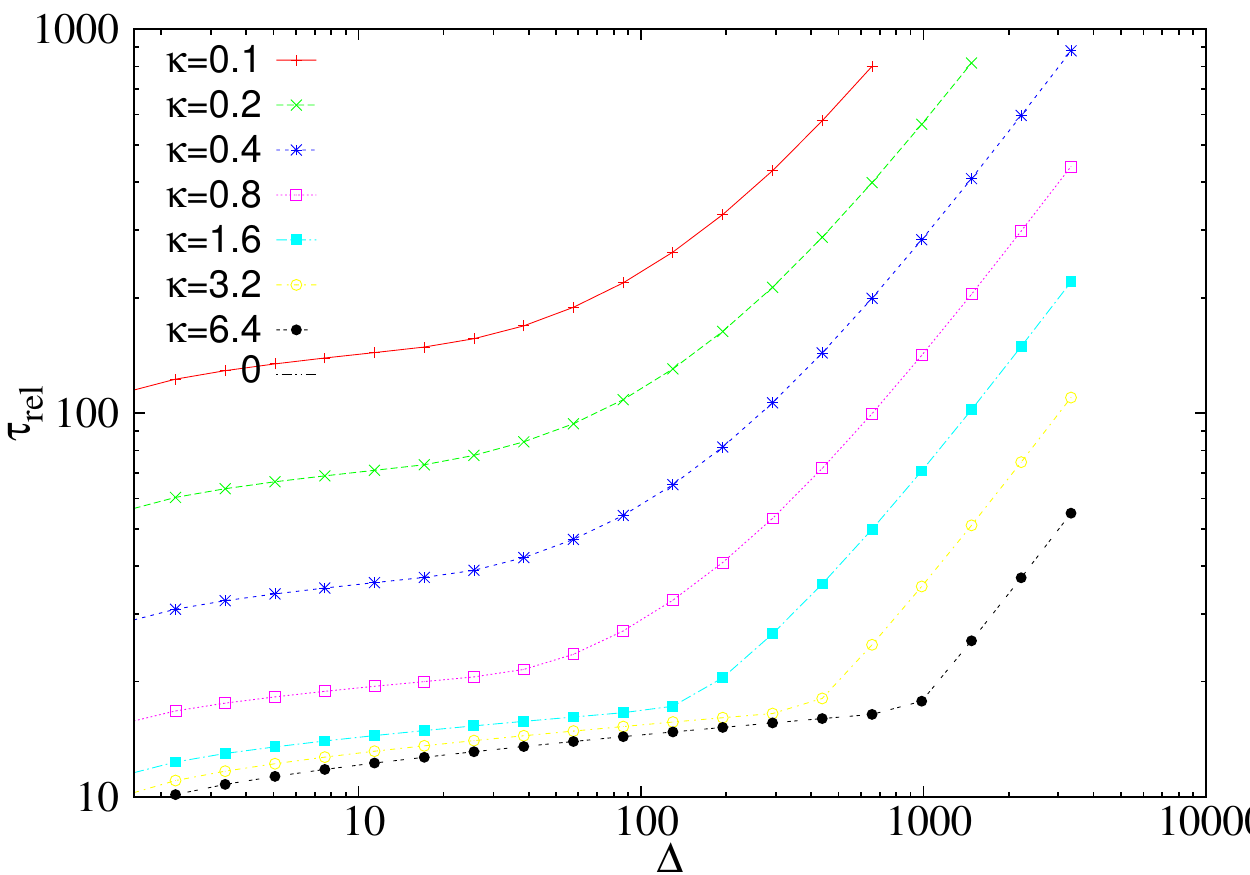}}
\caption{Relaxation time $\tau_{rel}$, as a function of the size of the perturbation, for different values of the rate of catalytic complex processing $\kappa$. Remaining kinetic parameters are as follows: $b_1=\beta=1$, $b_2=1$, $d_1=d_2=\delta=1$,$k^+_1=k^+_2=100$, $k_1^-=k_2^-=0$.
\label{fig5s}
}
\end{figure}

\begin{figure}
\centering
{\includegraphics[width=0.48\textwidth]{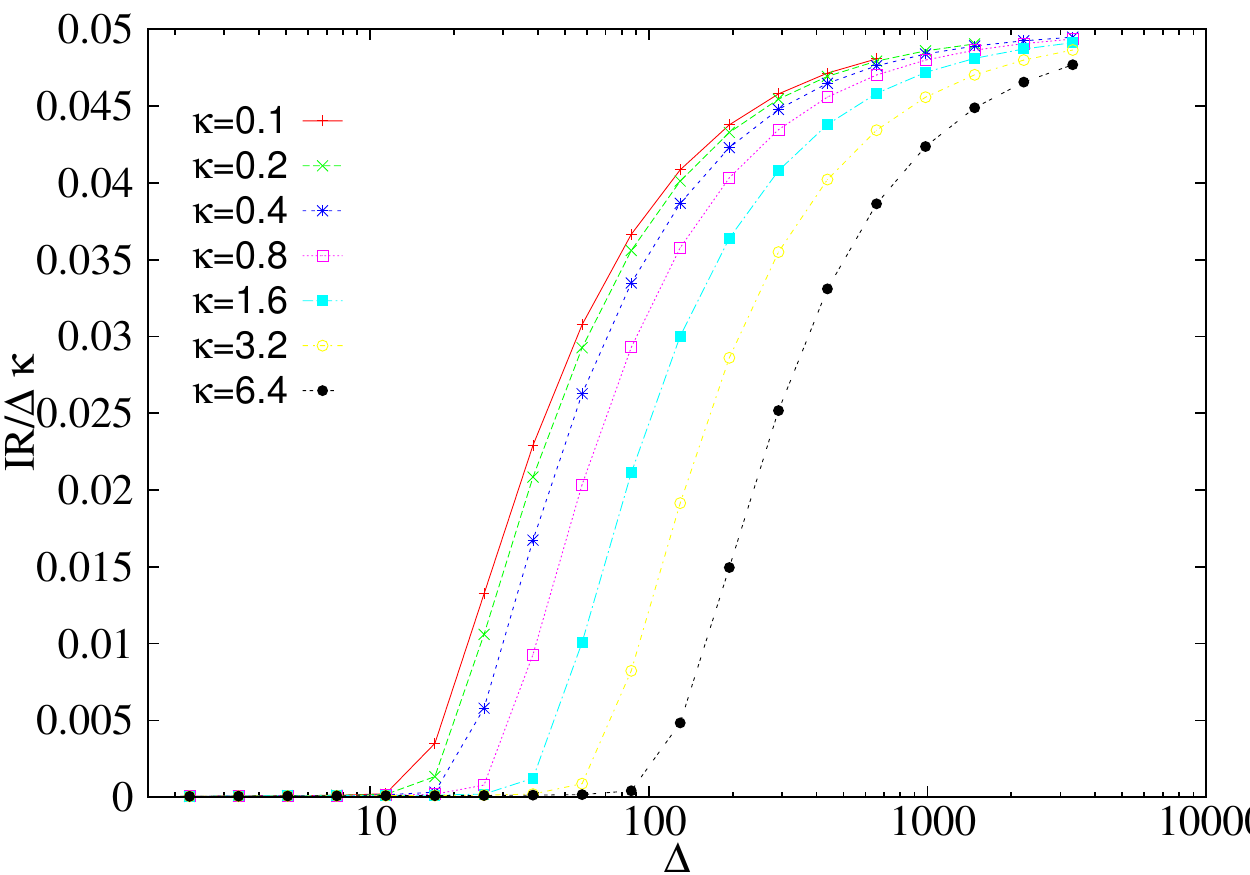}}
\caption{Integrated response IR between bound ceRNAs as a function of the perturbation size $\Delta$, for different processing rates $\kappa$. Remaining kinetic parameters are as follows: $b_1=b_2=1$, $b_2=1$, $\beta=1$, $d_1=d_2=\delta=1$, $k_1^-=k_2^-=0$, $\kappa_1=1$, $k_2^+=100$
\label{fig6s}
}
\end{figure}

Figures \ref{fig1s}, \ref{fig2s}, \ref{fig3s} and \ref{fig4s} show the dynamical susceptibility $\chi_{ij}(\omega)$ for pairs of `free', `susceptible' and `bound' ceRNAs in the different limit considered in the Main Text.

\subsection*{Estimate of the relaxation time following a large, saturating perturbation}

In the case of a kinetically homogeneous system, where binding is irreversible and remaining kinetic parameters are the same for all ceRNAs, in particular $d_i=d$, $k^+_i=k^+$, $k_i^-=k^-=0$, $\kappa_i=\kappa$ (and hence $\mu_{0,i}=\mu_{0}=\frac{d}{k^+}$) for all $i$, and assuming that ceRNAs and miRNAs reach a fast equilibrium with respect to the instantaneous values of the levels of the complexes, the following relations hold:

\begin{gather}
m_i(t)\simeq \frac{b_i}{d+k^+ \mu(t)} \quad i=1,...,N\\
\mu(t)\simeq \frac{\beta+\kappa\sum_j  c_j(t)}{\delta+ k^+ \sum_j m_j(t)}\\
\frac{dc_i(t)}{dt}=k^+ \mu(t) m_i(t)-\kappa c_i(t)  \quad i=1,...,N
\end{gather}
%with the intial condition $m_i(0)=m_i^\star, \mu= \beta/(\delta+\sum_k k^+ m_j^\star)$. 

If the perturbation is large enough, miRNAs are istantanously sequestered by the complexes and never undergo spontaneous decay, so that $k^+ \sum_j  m_j\gg\delta$.
%\begin{equation}
%\mu(t)\approx \frac{\beta+\sum_j (k^-+\kappa) c_j(t)}{\sum_j k^+ m_j(t)}
%\end{equation}
In this case one finds that the overall concentration of the complexes grows at constant rate $\beta$:
\begin{gather}
\sum_i \dot{c_i}=\beta
\end{gather}

It follows that:

\begin{gather}\label{large1}
\mu(t)\simeq \frac{\beta+\kappa\big(\sum_j  c_j(0)+ \beta t\big) }{\sum_j k^+ m_j(t)}\simeq \frac{\kappa\beta t}{k^+ \sum_j m_j(t)}
\end{gather}

for large enough $t$.

 The relaxation time $\tau_{rel}$ can be estimated by the condition
\begin{equation}\label{large2}
\mu(\tau_{rel}) \simeq \mu_{0}=\frac{\delta}{k^+}~~,
\end{equation}
or, accordingly,
\begin{equation}\label{large3}
m_i(\tau_{rel})\simeq \frac{m_i^\star}{2}~~.
\end{equation}
Plugging (\ref{large2}) and (\ref{large3}) in (\ref{large1}) one gets, in the limit of large perturbations $\Delta_i$:

\begin{equation}
\tau_{rel} \approx \frac{\mu_{0}k^+\sum_j m_j }{\beta \kappa} \approx \frac{\Delta_j b_j}{2\beta \kappa}
\end{equation}

where we have used:

\begin{equation}
\sum_i m_i(\tau_{rel})=\sum_i \frac{m_i^\star}{2}=\frac{(\sum_i b_i)+b_j\Delta_j}{2 \delta}\approx \frac{\Delta_jb_j}{2\delta}
\end{equation}

\clearpage

\end{document}